\documentclass[11pt,letterpaper]{article}

\usepackage{amssymb}
\usepackage{amsmath}
\usepackage{color}
\usepackage[pdftex]{graphicx}
\usepackage[margin=1in]{geometry}

\allowdisplaybreaks

\renewcommand{\baselinestretch}{1}
\newtheorem{cor}{Corollary}
\newtheorem{defin}{Definition}

\newtheorem{lem}{Lemma}
\newtheorem{prop}{Proposition}

\newtheorem{theo}{Theorem}

\begin{document}

\title{\textsc{Conditional Dominance in Games with Unawareness\thanks{I thank Martin Meier and Aviad Heifetz for their substantial input. The paper would not exist without them. Moreover, I am grateful to Pierpaolo Battigalli, Amanda Friedenberg, and Nicodemo De Vito for helpful discussions. I also thank the associate editor and three anonymous reviewers for very constructive comments that helped me to improve the paper. Financial support via ARO W911NF2210282 and NSF SES-0647811 is gratefully acknowledged.}}}

\author{Burkhard C. Schipper\thanks{Department of Economics, University of California, Davis. Email: bcschipper@ucdavis.edu}}

\date{Current Version: April 28, 2026; First Version: February 5, 2011}

\maketitle

\begin{abstract} Heifetz, Meier, and Schipper (2013) introduced dynamic games with unawareness consisting of a partially ordered set of games in extensive form. Here, we study the normal form of dynamic games with unawareness. The generalized normal form associated with a dynamic game with unawareness consists of a partially ordered set of games in normal form. We characterize strong rationalizability (resp., prudent rationalizability) in dynamic games with unawareness by iterated conditional strict (resp., weak) dominance in the associated generalized normal form. We also show that the analogue to iterated admissibility for dynamic games with unawareness depends on the extensive form. This is because, under unawareness, a player's information set not only determines which nodes she considers possible but also which game tree(s) she is aware of. \bigskip


\bigskip

\noindent \textbf{JEL-Classifications: } C72, D83.
\end{abstract}

\thispagestyle{empty}

\pagenumbering{empty}

\renewcommand{\baselinestretch}{1.1}

\small\normalsize

\pagenumbering{arabic}

\newpage

\section{Introduction}

Recent years have witnessed the extension of game theory to unawareness. The first challenge was to invent formal tools for modeling asymmetry in the players' (lack of) conception, beliefs about other players' lack of conception, etc. It goes beyond modeling standard asymmetric lack of information. Several frameworks have been proposed, both for static games (Meier and Schipper, 2014, Feinberg, 2021, Sadzik, 2021, Perea, 2022) and dynamic games (Halpern and Rego, 2014, Heifetz, Meier, and Schipper, 2013, 2021, Feinberg, 2021, Grant and Quiggin, 2013, Li, 2008, Schipper, 2021, Foo and Schipper, 2026, Guarino, 2020); see Schipper (2014) for a non-technical review. These tools already lead to interesting applications in contract theory (Filiz-Ozbay, 2012, von Thadden and Zhao, 2012, Auster, 2013, Auster and Pavoni, 2024, Francetich and Schipper, 2025), mechanism design (Pram and Schipper, 2025), disclosure of verifiable information under unawareness (Heifetz, Meier, and Schipper, 2021, Li and Schipper, 2025, Li, Peitz, and Zhao, 2016), political economy (Schipper and Woo, 2019), and financial market microstructure (Schipper and Zhou, 2021).\footnote{See http://www.econ.ucdavis.edu/faculty/schipper/unaw.htm for a bibliography on unawareness.} 

Developing sensible solution concepts for games with unawareness has been another challenge. Traditional concepts of equilibrium assume an ex-ante ready made behavioral convention for players. In games with unawareness, however, a player's awareness may change endogenously during the course of play making the equilibrium assumption implausible; see Schipper (2021) for an extensive discussion. As a remedy, rationalizability concepts have been proposed (Heifetz, Meier, and Schipper, 2013, 2021, Perea, 2022) and successfully used in applications (Schipper and Woo, 2019, Francetich and Schipper, 2025). It is well-known that rationalizability in the normal form is equivalent to the iterated elimination of strictly dominated strategies (Pearce, 1984). A similar result holds for strong rationalizability and iterated elimination of conditional strictly dominated strategies (Shimoji and Watson, 1998, Chen and Micali, 2013).\footnote{Strong rationalizability has also been called extensive-form rationalizability in the literature. It has been introduced by Pearce (1984) and Battigalli (1997). More specifically, we refer to what have been called extensive-form correlated rationalizability because for $n$-player games with $n > 2$ we allow any player to believe in correlations among her opponents' strategies.} Such a characterization is useful in applications, as it is often easier to eliminate strategies by dominance arguments than to eliminate beliefs over opponents' strategy profiles, especially in games with more than two players.\footnote{While finding both undominated strategies and best responses involve linear programming problems, the dimensionalities of these linear programming problems differ. For undominated strategies, we need to find mixtures over the player's \emph{own} strategies while for rationalizability, we need to find mixtures over opponents' strategy \emph{profiles}. With more than two players, the latter problem involves more computations.} This paper studies the connection between rationalizability and dominance in games with unawareness.

We study the equivalence between strong rationalizability and iterated conditional dominance in dynamic games with unawareness (Heifetz, Meier, and Schipper, 2013). Such an equivalence would be surprising in dynamic games with unawareness because the endogenous change in players' awareness during the game elevates the time structure of the strategic interaction. Iterated conditional dominance is applied to the associated normal form. Arguably, the normal form ``folds'' the time structure. Nevertheless, Kohlberg and Mertens (1986) argued that for standard games without unawareness, the normal form contains all strategically relevant information. Could a similar claim be made for strategic situations with unawareness in which players' perceptions of the situation change endogenously in ways that ex ante are not even anticipated by all players? To answer this question in the context of rationalizability notions for games in extensive form with unawareness, we first need to define the appropriate normal form associated with dynamic games with unawareness. For standard dynamic games without unawareness, the associated normal form features strategies of players as primitives. A player's strategy assigns an action to each of her information sets. Dynamic games with unawareness consist of a forest of partially ordered trees, each representing a partial view of the objectively feasible sequences of moves. A player's information set at a node of a tree may consist of nodes of a ``less expressive'' tree. Ex ante, a player may not be aware of all strategies and may discover further actions during the course of the play. Yet, for any tree, there is a well-defined set of partial strategies, namely assignments of actions to information sets of the player in the tree and in any less expressive trees. The associated generalized normal form consists now of a partially ordered set of normal forms, indexed by trees, taking the sets of \emph{partial} strategies as primitives.

Having defined the generalized normal form associated to dynamic games with unawareness, we can characterize strong rationalizability of Pearce (1984) and Battigalli (1997) extended to unawareness in Heifetz, Meier, and Schipper (2013) by iterated elimination of conditionally strictly dominated strategies. The characterization is similar to Shimoji and Watson (1998) for standard games without unawareness. Yet, the non-trivial twist is that partial strategies eliminated in less expressive normal forms must also trigger the elimination of strategies in more expressive normal forms that they are part of. This is because strategies in more expressive normal forms also specify actions in less expressive trees. A player with an awareness level given by the less expressive tree may consider the corresponding partial strategy to be dominated. Consequently, opponents with higher awareness levels realize that their rational but unaware opponent will not play such a strategy, thus eliminating this player's strategy in their more expressive normal form. We illustrate this novel feature with a simple example in the next section.

Strong rationalizability may involve imprudent behavior. It may be rationalizable for a player to make an opponent aware of one of the opponent's actions that is extremely harmful for the player because the player is allowed to believe that the opponent will not take this action. As a remedy, Heifetz, Meier, and Schipper (2021) introduce a version of strong rationalizability, called prudent rationalizability, using an idea of prudence or caution that proved to be instrumental in applications such as disclosure of verifiable information (Heifetz, Meier, and Schipper, 2013, Li and Schipper, 2020, 2025), electoral campaigning (Schipper and Woo, 2019), and screening under unawareness (Francetich and Schipper, 2025). Recently, De Vito (2025) introduced a rationalizability procedure using systems of conditional non-standard probability measures and a non-standard notion of strong belief (Battigalli and Siniscalchi, 2002). With these definitions, he is able to characterize prudent rationalizability as iterative reduction procedure on beliefs (in games without unawareness). In this paper, we characterize prudent rationalizability in dynamic games with unawareness by iterated elimination of conditionally weakly dominated strategies in the associated generalized normal form. Again, the twist is that (partial) strategies eliminated in less expressive normal forms must trigger also the elimination of any strategies in more expressive normal forms that they are part of. While games without unawareness are a special case of games with unawareness, and hence our result implies that prudent rationalizability is characterized by iterated admissibility also in games without unawareness, De Vito (2025) recently showed this for games without unawareness using his nice characterization of prudent rationalizability as a iterated reduction procedure on beliefs. 

One could argue that the characterization of strong rationalizability (resp., prudent rationalizability) by iterated conditional strict (resp., weak) dominance falls short of showing that sophisticated extensive-form reasoning can be captured in the normal form. This is because we condition on normal-form information sets corresponding to information sets of a \emph{given} dynamic game with unawareness. In our context, a relevant normal-form information set consists of the subset of partial strategy profiles inducing a path through an information set in a \emph{given} dynamic game with unawareness. Thus, we implicitly use extensive-form structures in the definition of iterated conditional strict (resp., weak) dominance when conditioning on normal-form information sets. In contrast, for standard games without unawareness, Shimoji and Watson (1998) show that sophisticated reasoning embodied in strong rationalizability can be captured by iterated elimination of conditionally strictly dominated strategies in the normal form using more generally any normal-form information set, i.e., normal-form information sets for which there exists \emph{some} game in extensive form with a corresponding information set and that normal form (Mailath, Samuelson, and Swinkels, 1993). Such a result is elusive for games with unawareness since normal-form information sets must also encode awareness that crucially depends on the extensive form. A potential remedy would be to use a different solution concept that does not use the extensive-form structure, even though it somehow captures sophisticated reasoning in the extensive form. Iterated admissibility is such a solution concept for standard games without unawareness. Indeed, Brandenburger and Friedenberg (2011) show that for standard dynamic games with perfect recall, iterated admissibility coincides with iterated conditional weak dominance at every level. Thus, by an inductive application of Lemma 4 of Pearce (1984), in standard games without unawareness, prudent rationalizability coincides with iterated admissibility in the associated normal form at every level/iteration. However, as we demonstrate, in dynamic games with unawareness the appropriate definition of iterated admissibility must make use of information sets as well. This is because information sets in dynamic games with unawareness not only model a player's information in the standard sense but also her awareness. The player's awareness of strategies is crucial for admissibility since, for instance, at the first level, a player cautiously considers possible any of the opponents' strategies only to the extent that she is aware of them. We define iterated admissibility for games with unawareness and show that in dynamic games with unawareness, iterated admissibility is conceptually closer to iterated conditional weak dominance because it cannot be independent of the awareness encoded in information sets of the extensive form. We show that prudent rationalizability is characterized by iterated admissibility in games with unawareness.

We should take the opportunity to clarify the connection between strong rationalizability and prudent rationalizability. Heifetz, Meier, and Schipper (2021) showed that prudent rationalizability is not a refinement of strong rationalizability in terms of strategies. However, they conjectured that it might be a refinement in terms of outcomes. Recently, Catonini (2024) showed that iterated admissibility is not an outcome refinement of strong rationalizability in standard dynamic games without unawareness. This is in contrast to static games, where iterated admissibility is known to be a refinement of rationalizability. Since our results imply that prudent rationalizability is equivalent to iterated admissibility, we conclude that prudent rationalizability is \emph{not} an outcome refinement of strong rationalizability in dynamic games. 

Pram and Schipper (2025) use insights from the current paper to study efficient implementation in \emph{conditional} weak dominant equilibrium in mechanisms under unawareness. In such settings, dynamic mechanisms become crucial as the awareness contained in the agents' reports is pooled by the mediator and communicated back to all agents who subsequently elaborate on their previously reported types at the pooled awareness level. They extend dominant strategy implementation of VCG mechanisms in the standard setting to \emph{conditional} dominant strategy implementation of dynamic direct elaboration VCG mechanisms under unawareness. They use the insight from the current paper that admissibility in games with unawareness must still condition on the awareness embodied in information sets. It necessitates the use of \emph{conditional} dominance under unawareness.

The paper is organized as follows: The next section provides an illustrative example. Section~\ref{GEFG} recalls definitions of dynamic games with unawareness and presents the novel notion of associated generalized normal form. Section~\ref{EFR} shows the characterization of strong rationalizability by iterated elimination of conditional strictly dominated strategies. The characterization of prudent rationalizability by iterated elimination of conditional weakly dominated strategies is presented in Section~\ref{PR}. Finally, we present the notion of iterated admissibility and show that it characterizes prudent rationalizability in Section~\ref{IA}. Proofs are relegated to an appendix.

\section{Introductory Example~\label{Example}}

\begin{figure}[!h]
\caption{Example without Unawareness\label{standard}}
\begin{center}
\includegraphics[scale=.65]{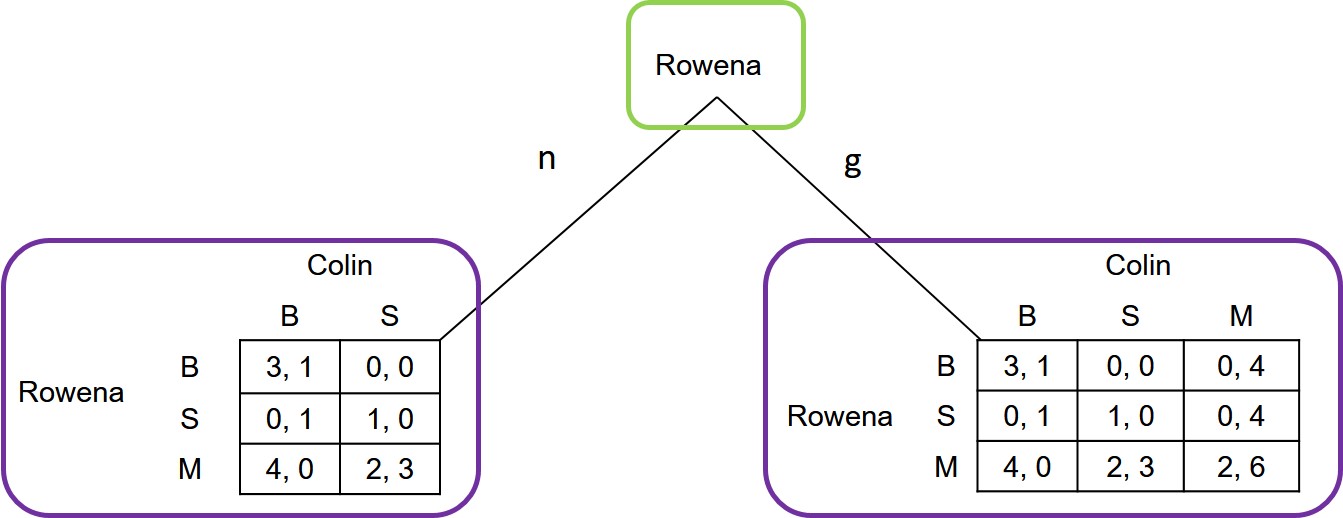}
\end{center}
\end{figure}

The following example aims to illustrate (a) the novel notion of generalized normal form associated with dynamic games with unawareness and (b) the subtleties of the characterization of strong rationalizability by iterated conditional dominance in games with unawareness. In particular, we contrast it to standard games. We begin with a standard game without unawareness (see Figure~\ref{standard}) between two players, Rowena and Colin.
Rowena moves first, deciding between actions $n$ and $g$. This is followed by a simultaneous-move game. The implication of Rowena's action $g$ is that Colin has available all three actions $B$, $S$, and $M$. Otherwise, if Rowena takes action $n$, Colin is just left with actions $B$ and $S$ in the simultaneous game that follows. That is, Rowena's initial move affects Colin's availability of action $M$.\footnote{While the names of actions and the structure bears some similarity with a game discussed in Heifetz, Meier, and Schipper (2013), the payoffs differ.} The information sets of players are indicated by the green and purple rectangles. The green rectangle at the root of the tree belongs to Rowena. The purple rectangles indicate information sets of both players.

Strong rationalizability is almost trivial in this example, as the procedure concludes after the first level. Any strategy of Rowena that does not prescribe action $M$ after action $n$ is not strongly rationalizable, as there is no belief over Colin's strategies conditional on reaching the subgame after $n$ for which any action other than $M$ is rational. Similarly, any strategy of Rowena that does not prescribe action $M$ after action $g$ is not strongly rationalizable. This leaves strategies $\{nMB, nMS, nMM, gBM, gSM, gMM\}$ for Rowena (where the first component refers to Rowena's action at the root of the tree, the second to her action in the left subgame, and the last component to her action in the right subgame). For Colin, any strategy that does not prescribe $M$ in the subgame after $g$ is not strongly rationalizable since there is no belief of Colin over Rowena's strategies conditional on $g$ that would make actions $B$ or $S$ rational in the subgame after $g$. In contrast, $B$ can be rationalized by Colin after $n$ with a belief that puts sufficiently high probability on Rowena playing $B$ after $n$. This leaves strategies $\{BM, SM\}$ for Colin. At the second level, both players believe in the opponent's first-level strongly rationalizable strategies. In this example, strong rationalizability does not lead to further eliminations. However, the example is sufficiently rich to demonstrate the difference between strict dominance and \emph{conditional} strict dominance.
\begin{figure}[!h]
\caption{Associated Normal Form\label{standardNF}}
\begin{center}
\includegraphics[scale=.13]{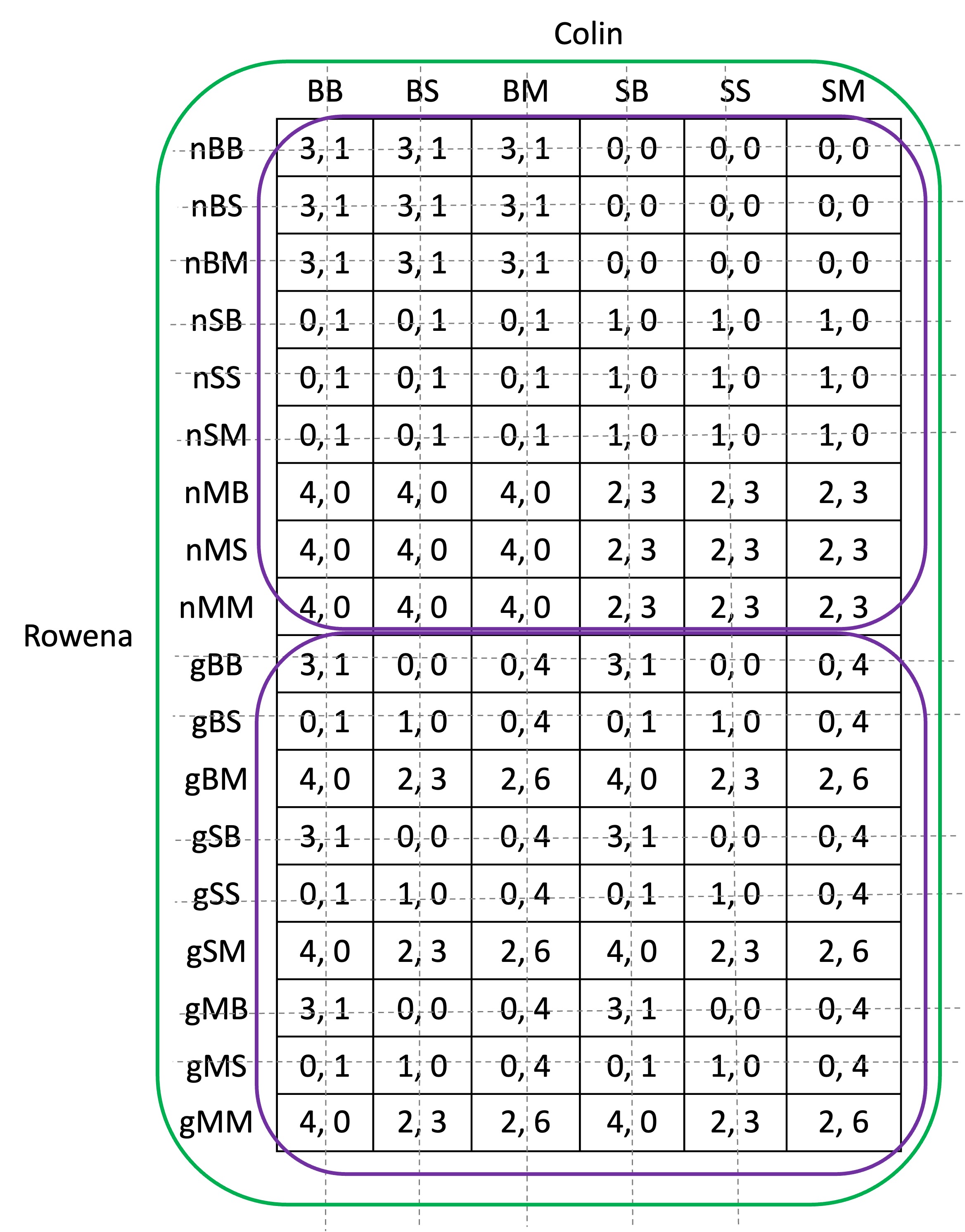}
\end{center}
\end{figure}

Figure~\ref{standardNF} presents the normal form associated with the standard dynamic game of Figure~\ref{standard}. The rows represent Rowena's strategies. E.g., $nBS$ denotes the strategy that prescribes $n$ at her initial information set, $B$ at her information set after $n$, and $S$ at her information set after $g$. The strategies of Colin are represented by columns. E.g., $BS$ is the strategy that would require Colin to play $B$ at his information set after $n$ and $S$ at his information set after $g$. We also indicate the normal-form information sets by rectangles. A normal-form information set consists of the subset of strategy profiles inducing a path through an information set in the dynamic game. E.g., the entire strategy space is consistent with Rowena's first information set. That is, the green information set in Figure~\ref{standard} corresponds to the green rectangle in Figure~\ref{standardNF}.

We indicate strategies eliminated by conditional strict dominance by dashed lines. Notably, strategies $\{BB, BS, SB, SS\}$ of Colin are eliminated because they are strictly dominated \emph{conditional} on the normal-form information set at the bottom of the figure, which corresponds to his information set after $g$. These strategies would not be \emph{unconditionally} strictly dominated, which illustrates how conditional strict dominance goes beyond (unconditional) strict dominance. We also observe that for both players, strong rationalizable strategies coincide exactly with the strategies obtained by elimination of strictly conditional dominated strategies. This is just an example of the general result by Shimoji and Watson (1998).
\begin{figure}[!h]
\caption{Example of a Dynamic Game with Unawareness\label{modBoS}}
\begin{center}
\includegraphics[scale=0.65]{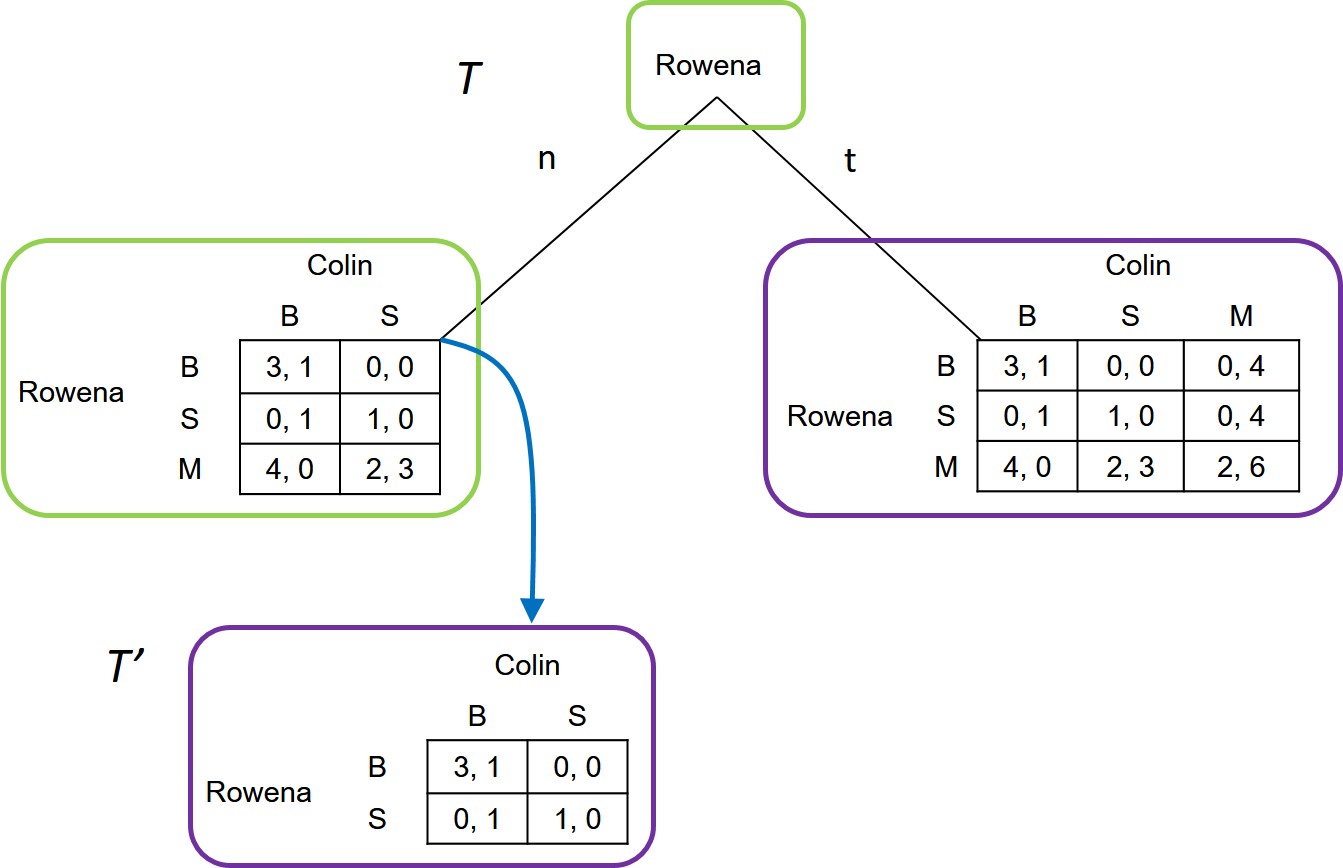}
\end{center}
\end{figure}

Consider now an analogous game in which Colin is unaware of action $M$ (rather than not having action $M$ available) depicted in Figure~\ref{modBoS}. There are two ``trees'', the upper tree $T$ and the lower ``tree'' $T'$. The upper tree looks equivalent to the tree in Figure~\ref{standard}. However, Rowena's action $t$ is interpreted as raising Colin's awareness of action $M$. This is indicated by the purple information after $t$, an information set that belongs to both players. If Rowena takes action $n$ instead, then Colin remains unaware of action $M$. This is indicated by his information set after $n$, which is not on tree $T$ but on the lower ``tree'' $T'$ (see the blue arrow). The lower ``tree'' is just a simultaneous move game between Rowena and Colin with actions $B$ and $S$. There is no mentioning of $M$. Even action $n$ of not telling Colin about $M$ is not part of the description of the strategic situation in the lower tree $T'$, indicating that when a player considers the tree $T'$, (s)he is unaware of $M$ (and of $n$ and $t$). The green information set after action $n$ belongs only to Rowena. Colin's information set after $n$ is the purple information at $T'$. This purple information set is also the information set of both players in the lower ``partial'' game. The example in Figure~\ref{modBoS} illustrates two important features of dynamic games with unawareness (Heifetz, Meier, and Schipper, 2013). First, there is an ordered set of trees. In the example, $T$ is more expressive than $T'$. Second, an information set of a player after a history in one tree may consist of histories in a less expressive tree. The latter feature models unawareness of that player of some action in the tree that is not represented in the less expressive tree. Colin's information set after $n$ in $T$ consists only of nodes of the less expressive tree $T'$. 

In this example, strong rationalizability concludes after two rounds/levels. At the first level, any strategy of Rowena that does not prescribe $M$ after $n$ is not strongly rationalizable because there is no belief over Colin's strategies conditional on $n$ that would let $B$ and $S$ yield a higher expected payoff than $M$ after $n$. An analogous argument holds for strategies with initial action $t$. Thus, at the first level, only strategies in the set $\{ n M * *, t * M *\}$ are strongly rationalizable for Rowena, where the first component refers to her actions $n$ and $t$ at her initial information set, the second component refers to her action in the simultaneous move game after $n$, the third component specifies her action in the simultaneous game after $t$, and the fourth component belongs to the simultaneous move game making up the lower ``tree'' $T'$. (We indicate by ``$*$'' that the action at the corresponding information set can be arbitrary.) For Colin, the only strongly rationalizable strategy at the first level is $MB$, where the first component refers to his action at his information set in $T$ after she takes $t$ and the second component specifies his action at his information set in ``tree'' $T'$.

At the second level, both players can take into account first-level strongly rationalizable strategies of their opponent. Hence, upon reaching the information set after $t$, Rowena is certain that Colin plays $M$, whereas when she selects $n$, Colin's action is $B$ because he remains unaware of $M$ and perceives the strategic situation to be given by $T'$. Colin's action $B$ in $T'$ induces the action $B$ in the simultaneous move game after $n$. Thus, with any such belief, Rowena's expected payoff is largest when she plays a strategy with $n$ followed by $M$. Also, in the lower tree $T'$, the unaware incarnation of Rowena is certain that Colin plays $B$. Thus, her best response is $B$ in $T'$. Hence, the set of second-level strongly rationalizable strategies of Rowena is $\{n M * B\}$. For Colin, it remains $\{MB\}$. No further changes occur at higher levels.

\begin{figure}[!hp]
\caption{Associated Generalized Normal-form Game of the Example\label{modBoSNF}}
\begin{center}
\includegraphics[scale=0.72]{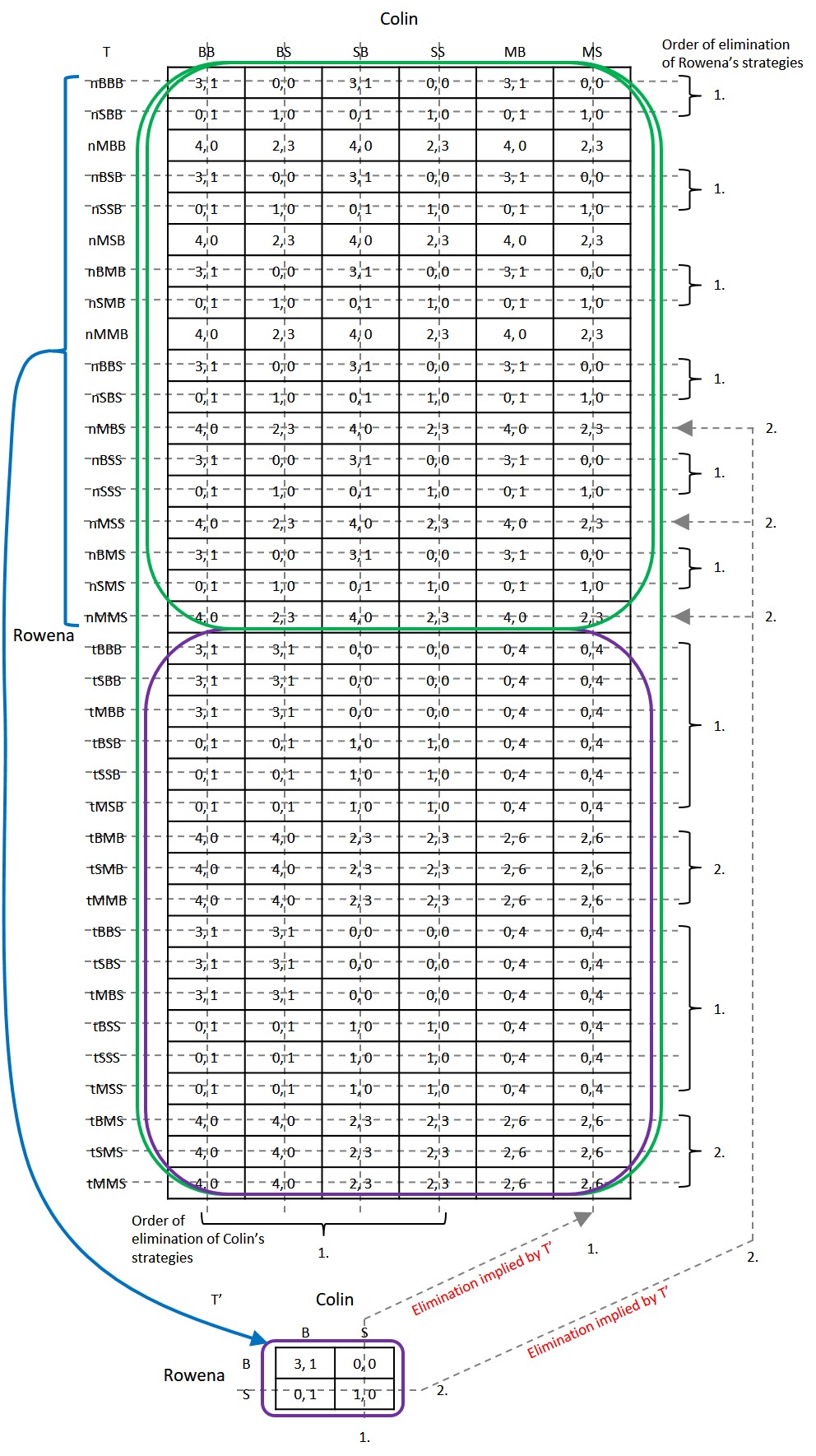}
\end{center}
\end{figure}

To derive the strategies surviving iterated elimination of strictly dominated strategies, we first need to consider the associated normal form. For standard dynamic games, the associated game in normal form is the game in which the players' strategies of the dynamic game are the primitives. A strategy assigns an action to each information set of a player. Since in a standard dynamic game every player is aware of all actions, in principle she can ``control'' her entire strategy ex ante. In dynamic games with unawareness, for each information set of a player, her strategy specifies -- from the point of view of the modeler -- what the player would do if and when that information set of hers is ever reached. In this sense, a player does not necessarily ``own'' her full strategy at the beginning of the game, because she might not be initially aware of all her information sets. For instance, in Figure~\ref{modBoS} a strategy of Colin assigns an action to the information set in the upper tree $T$ and an action to the information set in the lower tree $T'$. A strategy for Rowena assigns an action to the root of the upper tree $T$, the left information set in upper tree, the right information set in upper tree, and the information set in the lower tree $T'$. When players face the game in the lower tree $T'$, they can only choose partial strategies, i.e., strategies restricted to the information set in $T'$ because they are unaware of the upper tree $T$. The set of $T'$-partial strategies is just $\{B, S\}$ for any player. With the notion of strategy and partial strategy, we can define the associated generalized normal form as usual by taking the strategies and partial strategies as primitives. When taking partial strategies in $T'$ as primitives, we get the normal form at the bottom of Figure~\ref{modBoSNF}. When we take entire strategies as primitives, we obtain the upper normal form in Figure~\ref{modBoSNF}. That is, while a dynamic game with unawareness consists of a partially ordered set of trees, its generalized associated normal form consists of a partially ordered set of normal forms. The operation of deriving the associated generalized normal form from a game in extensive form with unawareness is analogous to standard games.

We also indicated the normal-form information sets in Figure~\ref{modBoSNF} using the same color coding as for the corresponding information sets in Figure~\ref{modBoS}. Green belongs to Rowena only, purple to both players, and the blue arrow indicates that when Rowena chooses any strategy with action $n$, Colin remains unaware of $M$ and his information set is in the lower tree $T'$. Here, the normal-form information set after Rowena chooses action $n$ in $T$ is the entire normal form associated with the lower tree $T'$.

With the dashed lines and the numbers beside them, we indicate the order of iterated elimination of conditional strictly dominated strategies. However, in games with unawareness the algorithm becomes more subtle since conditional dominance of a $T'$-partial strategy triggers deletion of all strategies in the upper normal form that feature this partial strategy as a component. This is the case for Colin in the first round, where the deletion of $S$ in the game $T'$ implies that all strategies with $S$ as the second component in the game $T$ are eliminated as well. In particular, this applies to strategy $MS$ that is not otherwise conditionally strictly dominated in the upper normal form associated with $T$.\footnote{Again, we emphasize that while our example discussed here bears similarities with examples discussed in Heifetz, Meier, and Schipper (2013), the payoffs differ. The reason is two-fold: First, we want illustrate that a partial strategy, which is conditionally dominated in a lower normal form, triggers deletion of all strategies in the upper normal form that feature the partial strategy as a component. Second, while in the example of Heifetz, Meier, and Schipper (2013), strong rationalizability in a game with unavailability of an action yields a sharper prediction than in an analogous game with unawareness of that action, the opposite is the case in the example discussed here. We conclude that strong rationalizability/iterated conditional strict dominance with unawareness of actions differs from unavailability of actions but none necessarily yields a sharper prediction than the other.} We indicate this with dashed arrows. When considering Colin's play, Rowena anticipates in $T$ that Colin will not play $MS$ because $S$ is strictly dominated for him in $T'$. After two rounds, the process ends. The remaining strategies coincide exactly with the strong rationalizable strategies. More precisely, the strategies that remain after each round of elimination of conditionally strictly dominated strategies are exactly those strategies that remain after the corresponding round of the strong rationalizability procedure. In Section~\ref{EFR}, we show that this is generally the case.

\section{Dynamic Games with Unawareness\label{GEFG}}

In this section we outline dynamic games with unawareness as introduced by Heifetz, Meier, and Schipper (2013).\footnote{See also Heifetz, Meier, and Schipper (2021), Schipper (2021), and Foo and Schipper (2026).} To define a dynamic game $\Gamma$, consider first, as a building block, a finite perfect information game with a set of players $I,$ a set of decision nodes $N_{0},$ active players $I_{n}$ at node $n$ with finite action sets $A_{n}^{i}$ of player $i\in I_{n}$ (for $n \in N_{0}$), and terminal nodes $Z_{0}$ with a payoff vector $\left( p_{i}^{z}\right) _{i\in I}\in \mathbb{R}^{I}$ for the players for every $z\in Z_{0}$. The nodes $\bar{N}_{0}=N_{0} \cup Z_{0}$ constitute a tree.\footnote{Heifetz, Meier, and Schipper (2013) considered also moves of nature which we do not explicitly consider here. Note that we could treat nature just like another player who is indifferent among all terminal nodes.}

Consider now a family $\mathbf{T}$ of subtrees of $\bar{N}_{0}$, partially ordered ($\preceq $) by inclusion. One of the trees $T_{1} \in \mathbf{T}$ is meant to represent the modeler's view of the paths of play that are \emph{objectively} feasible; each other tree represents the feasible paths of play as \emph{subjectively} viewed by some player at some node at one of the trees.

In each tree $T \in \mathbf{T}$, denote by $n_{T}$ the copy in $T$ of the node $n \in \bar{N}_0$ whenever the copy of $n$ is part of the tree $T$. However, in what follows we will typically avoid the subscript $T$ when no confusion can arise.

Denote by $N_{i}^{T}$ the set of nodes in which player $i \in I$ is active in the tree $T \in \mathbf{T}$. We require that all the terminal nodes in each tree $T\in \mathbf{T}$ are copies of nodes in $Z_{0}$. Moreover, if for two decision nodes $n$, $n^{\prime }\in N_{i}^{T}$ (i.e., $i\in I_{n}\cap I_{n^{\prime }}$) it is the case that $A_{n}^{i}\cap A_{n^{\prime }}^{i}\neq \emptyset $, then $A_{n}^{i}=A_{n^{\prime }}^{i}$.

Denote by $N$ the union of all decision nodes in all trees $T \in \mathbf{T}$, by $Z$ the union of terminal nodes, and by $\bar{N}=N \cup Z$. (Copies $n_T$ of a given node $n$ in different subtrees $T$ are distinct from one another, so that $\bar{N}$ is a disjoint union of sets of nodes.) For a node $n \in \bar{N}$ we denote by $T_{n}$ the tree containing $n$.

For each decision node $n \in N$ and each active player $i \in I_{n}$, the information set is denoted by $\pi_{i}\left( n\right)$. It is the set of nodes that the player $i$ considers as possible at $n$. The information set $\pi_{i}\left( n \right)$ will be in a tree different from tree $T_{n}$ if at $n$ the player is unaware of some of the paths in $T_{n}$, and rather envisages the dynamic interaction as taking place in the tree containing $\pi_{i}\left( n \right)$. We impose properties analogous to standard dynamic games with perfect recall; see Appendix~\ref{properties} and discussions in Heifetz, Meier, and Schipper (2013), Schipper (2021), and Foo and Schipper (2026).

We denote by $H_{i}$ the set of $i$'s information sets in all trees. For any information set $h_{i}\in H_{i},$ we denote by $T_{h_{i}}$ the tree containing $h_{i}.$ For two information sets $h_{i},h_{i}^{\prime }$ in a given tree $T,$ we say that $h_{i}$ precedes $h_{i}^{\prime }$ (or that $h_{i}^{\prime }$ succeeds $h_{i}$) if for every $n^{\prime }\in h_{i}^{\prime }$ \ there is a path $n,...,n^{\prime }$ such that $n\in h_{i}$. We denote the precedence relation by $h_{i}\rightsquigarrow h_{i}^{\prime }$.

Standard properties imply that if $n^{\prime}, n^{\prime \prime }\in h_{i}$ where $h_{i}=\pi _{i}\left(n\right) $ is an information set, then $A_{n^{\prime }}^{i}=A_{n^{\prime \prime }}^{i}$ (see Heifetz, Meier, and Schipper, 2013, Remark 1, for details). Thus, if $n \in h_i$ we write also $A_{h_i}$ for $A^i_{n}$.

Perfect recall guarantees that with the precedence relation $\rightsquigarrow $ player $i$'s information sets $H_{i}$ form an \emph{arborescence}: For every information set $h_{i}^{\prime }\in H_{i}$, the information sets preceding it $\left\{ h_{i}\in H_{i}:h_{i}\rightsquigarrow h_{i}^{\prime }\right\} $ are totally ordered by $\rightsquigarrow$.

For trees $T,T^{\prime }\in \mathbf{T}$ we denote $T \rightarrowtail T^{\prime}$ whenever for some node $n\in T$ and some player $i\in I_{n}$ it is the case that $\pi _{i}\left( n\right) \subseteq T^{\prime }$. Denote by $\hookrightarrow $\ the transitive closure of $\rightarrowtail$. That is, $T\hookrightarrow T^{\prime \prime }$ if and only if there is a sequence of trees $T,T^{\prime },\dots ,T^{\prime \prime }\in \mathbf{T}$ satisfying $T\rightarrowtail T^{\prime }\rightarrowtail \dots \rightarrowtail T^{\prime \prime }$. In the following, we assume that $T_1\hookrightarrow T^{}$, for all $T \in \mathbf{T}$, with $T \neq T_1$.

A dynamic game with unawareness $\Gamma $ consists of a partially ordered set $\mathbf{T}$ of subtrees of a tree $\bar{N}_{0}$ along with information sets $\pi_{i}\left(n\right)$ for every $n\in T$, $T\in \mathbf{T}$ and $i\in I_{n}$, satisfying all properties imposed by Heifetz, Meier, and Schipper (2013); see Appendix~\ref{properties}.

For every tree $T\in \mathbf{T}$, the $T$\emph{-partial game} is the partially ordered set of trees including $T$ and all trees $T'$ in $\Gamma $ satisfying $T \hookrightarrow T'$, with information sets as defined in $\Gamma $. A $T$-partial game is a dynamic game with unawareness, i.e., it satisfies the same properties.

We denote by $H_{i}^{T}$ the set of $i$'s information sets in the $T$-partial game.

A (pure) strategy
\begin{equation*}
s_{i} \in S_{i} \equiv \prod_{h_{i} \in H_{i}} A_{h_{i}}
\end{equation*}
for player $i$ specifies an action of player $i$ at each of her information
sets $h_{i}\in H_{i}$. Denote by
\begin{equation*}
S = \prod_{j \in I} S_{j}
\end{equation*}
the set of strategy profiles in the dynamic game with unawareness.

If $s_{i} = \left(a_{h_{i}}\right) _{h_{i}\in H_{i}}\in S_{i}$,
we denote by
\begin{equation*}
s_{i}\left( h_{i}\right) =a_{h_{i}}
\end{equation*}
the player's action at the information set $h_{i}$. If player $i$ is active at node $n,$ we say that at node $n$ the strategy prescribes to her the action $s_{i}\left( \pi _{i}\left( n \right) \right)$.

In dynamic games with unawareness, a strategy cannot be conceived as an ex ante plan of actions, one for each information set. If $h_{i}\subseteq T$ but $T \nsucceq T^{\prime }$, then at $h_{i}$ player $i$ may be interpreted as being unaware of her information sets in $H_{i}^{T^{\prime }} \setminus H_i^T$. Thus, a strategy of player $i$ should rather be viewed as a list of answers to the hypothetical questions \textquotedblleft what would the player do if $h_{i}$ were the set of nodes she considered as possible?\textquotedblright , for $h_{i}\in H_{i}.$ However, there is no guarantee that such a question about the information set $h_{i}^{\prime }\in H_{i}^{T^{\prime }}$ would even be meaningful to the player if it were asked at a different information set $h_{i}\in H_{i}^{T}$ when $T\not\hookrightarrow T^{\prime }$. The answer should therefore be interpreted as given by the modeler, as part of the description of the situation.

For a strategy $s_{i}\in S_{i}$ and a tree $T\in \mathbf{T}$, we denote by $s_{i}^{T}$ the strategy in the $T$-partial game induced by $s_{i}$. If $R_{i}\subseteq S_{i}$ is a set of strategies of player $i$, denote by $R_{i}^{T}$ the set of strategies induced by $R_{i}$ in the $T$-partial game. The set of $i$'s strategies in the $T$-partial game is thus denoted by $S_{i}^{T}$. Denote by $S^{T} = \prod_{j \in I}S_{j}^{T}$ the set of strategy profiles in the $T$-partial game. Note that, by our assumption, we have  $S_{i}^{T_1} = S_i$, for $i \in I$. 

We say that a strategy profile $s \in S$ \emph{reaches} the information set $h_{i}\in H_{i}$ if the players' actions in $T_{h_{i}}$ lead to $h_{i}$. (Notice that unlike in standard games, an information set $\pi_{i}(n) $ may be contained in tree $T^{\prime }\neq T_{n}$. In such a case, $s_{i}\left(\pi_{i}\left( n\right) \right)$ induces an action for player $i$ also in $n$ and not only at the nodes of $\pi_{i}(n)$.) We say that the strategy $s_{i}\in S_{i}$ \emph{allows} for the information set $h_{i}$ if there is a strategy profile $s_{-i}\in S_{-i}$ of the other players such that the strategy profile $\left( s_{i}, s_{-i}\right)$ reaches $h_{i}$. Otherwise, we say that the information set $h_{i}$ is excluded by the strategy $s_{i}$. Similarly, we say that the profile of strategies of player $i$'s opponents $s_{-i}\in S_{-i}$ \emph{allows} for the information set $h_{i}$ if there exists a strategy $s_{i}\in S_{i}$ such that the strategy profile $\left(s_{i}, s_{-i}\right)$ reaches $h_{i}$. A strategy profile $\left( s_{j}\right)_{j\in I}$ \emph{reaches a node} $n \in T$ if the players' actions $s_{j}\left( \pi _{j}\left( n^{\prime} \right) \right) _{j\in I}$ lead to $n$. Since we consider only finite trees, $\left( s_{j}\right) _{j\in I}$ reaches an information set $h_{i}\in H_{i}$ if and only if there is a node $n\in h_{i}$ such that $\left( s_{j}\right) _{j\in I}$ reaches $n$.

Like in standard games, a profile of strategies of the players induces exactly one terminal node in each tree, and hence defines a payoff for each player in each tree.

For an information set $h_{i},$ let $s_{i} / \tilde{s}_{i}^{h_{i}}$ denote the strategy that is obtained by replacing actions prescribed by $s_{i}$ at the information set $h_{i}$ and its successors by actions prescribed by $\tilde{s}_{i}$. The strategy $s_{i}/\tilde{s}_{i}^{h_{i}}$ is called an $h_{i}$-replacement of $s_{i}$.

\subsection{Associated Generalized Normal Form}

Consider a dynamic game with unawareness $\Gamma $ with a partially ordered set of trees $\mathbf{T}$. The \emph{associated generalized normal form} $G$ is defined by $\langle I, \langle (S_{i}^{T})_{i \in I},(u_{i}^{T})_{i\in I} \rangle_{T\in \mathbf{T}} \rangle$, where $I$ is the set of players in $\Gamma $ and $S_{i}^{T}$ is player $i$'s set of $T$-partial strategies. Recall that if player $i$ is active at node $n \in T$, then the strategy $s_{i}\in S_{i}^{T}$ assigns the action $s_{i}(\pi_{i}(n))$ to node $n$. Hence, each profile of strategies in $S^{T}$ induces a terminal node in $T$ (even if there is a player active in $T$ with no information set in $T$). We let $u_{i}^{T}(s)$ be the payoff associated with the terminal node in $T$ reached by $s \in S^{T}$. (Note that while strategy profiles in $S^{T}$ reach terminal nodes also in trees $T^{\prime }\in \mathbf{T}$, $T \hookrightarrow T^{\prime }$, $u_{i}^{T}$ concerns payoffs in the tree $T$ only.)

Recall that $H_{i}^{T}$ denotes player $i$'s set of information sets in the $T$-partial dynamic game. For each $h_{i}\in H_{i}^{T}$, let $S^{T}(h_{i})\subseteq S^{T}$ be the subset of the $T$-partial strategy space containing $T$-partial strategy profiles that reach the information set $h_{i}$. Define also $S_{i}^{T}\left( h_{i}\right) \subseteq S_{i}^{T}$ and $S_{-i}^{T}\left( h_{i}\right) \subseteq S_{-i}^{T}$ to be the set of player $i$'s $T$-partial strategies allowing for $h_{i}$ and the set of profiles of the other players' $T$-partial strategies allowing for $h_{i}$, respectively. For the entire game denote by $S(h_{i})\subseteq S$ the set of strategy profiles that reach $h_{i}$. Similarly, $S_{i}\left( h_{i}\right) \subseteq S_{i}$ is the set of player $i$'s strategies allowing for $h_{i}$ and $S_{-i}\left( h_{i}\right) \subseteq S_{-i}$ is the set of profiles of the other players' strategies allowing for $h_{i}$.

Given $\Gamma$ and its associated generalized normal form $G$, define player $i$'s set of \emph{normal-form information sets} by
\begin{equation*}
\mathcal{X}_{i}=\{S^{T_{h_i}}(h_{i}):h_{i}\in H_{i}\}.
\end{equation*}
These are the ``normal-form versions'' of information sets in the dynamic game with unawareness. In the literature on standard games, normal-form information sets refer more generally to subsets of the strategy space of a pure strategy reduced normal form for which there exists a game in extensive form with that normal form and corresponding information sets (see Mailath, Samuelson, and Swinkels, 1993). For our characterization, we are just interested in the normal-form versions of information sets of a \emph{given} dynamic game with unawareness.

For $T \in \mathbf{T}$, any set $Y\subseteq S^T$ is called a \emph{restriction for player $i$} (or an $i$-product set) of $T$-partial strategies if $Y = Y_{i} \times Y_{-i}$ for some $Y_{i}\subseteq S^T_{i}$ and $Y_{-i}\subseteq S^T_{-i}$. Clearly, a player's normal-form information set is a restriction for that player. That is, if $S^{T_{h_i}}(h_i)$ is a normal-form information set of player $i$, then it is a restriction for player $i$ of $T_{h_i}$-partial strategy profiles. We say that $Y \subseteq S^T$ is a \emph{restriction} of $T$-partial strategies if it is a restriction for every player.

\section{Strong Rationalizability and Iterated Conditional Strict Dominance\label{EFR}}

Strong rationalizability is an iterative procedure of refining players' beliefs about opponents' (partial) strategies. For each round $\ell$ of the procedure, each player forms beliefs at each of her information sets about her opponents' strongly rationalizable (partial) $\ell - 1$-level strategies consistent with the information set. She keeps only strategies of her own that are rational w.r.t. such systems of beliefs; this yields her set of strongly rationalizable $\ell$-level strategies.

In order to define the solution concept, we first define belief systems. A \emph{belief system} of player $i \in I$,
\begin{equation*}
b_{i}=\left( b_{i}\left( h_{i}\right) \right) _{h_{i}\in H_{i}}\in
\prod_{h_{i}\in H_{i}}\Delta \left( S_{-i}^{T_{h_{i}}}\right)
\end{equation*}
is a profile of beliefs, a belief $b_{i}\left( h_{i}\right) \in \Delta
\left( S_{-i}^{T_{h_{i}}}\right)$ about the other players' strategies in
the $T_{h_{i}}$-partial game, for each information set $h_{i}\in H_{i}$,
with the following properties:
\begin{itemize}
\item The belief $b_{i}\left( h_{i}\right)$ assigns probability 1 to the set of strategy profiles of the other players in the $T_{h_i}$-partial game that allow for $h_{i}$.

\item If $h_{i}$ precedes $h_{i}^{\prime }$ ($h_{i}\rightsquigarrow h_{i}^{\prime }$), then $b_{i}\left( h_{i}^{\prime} \right)$ is derived from $b_{i}\left( h_{i}\right)$ by conditioning whenever possible.\footnote{Note that by the definition of the precedence relation, $h_i$ and $h'_i$ belong to the same tree.} 
\end{itemize}
Denote by $B_{i}$ the set of player $i$'s belief systems.

Given a belief system $b_{i}\in B_{i}$, a strategy $s_{i}\in S_{i}$, and an information set $h_{i}\in H_{i}$, define player $i$'s expected payoff at $h_{i}$ as the expected payoff for player $i$ in $T_{h_{i}}$ given $b_{i}\left(h_{i}\right)$, the actions prescribed by $s_{i}$ at $h_{i}$ and its successors, and conditional on the fact that $h_{i}$ has been reached.\footnote{Even if this condition is counterfactual due to the fact that the strategy $s_{i}$ does not allow for $h_{i}$. The conditioning is thus on the event that player $i$'s past actions (at the information sets preceding $h_{i}$) have led to $h_{i}$ \emph{even if these actions are distinct from those prescribed by} $s_{i}$.}

We say that given a belief system $b_{i}$, the strategy $s_{i}$ is \emph{rational} at the information set $h_{i}\in H_{i}$ for $b_i$, if either $s_{i} $ does not allow for $h_{i}$ in the tree $T_{h_{i}}$, or if $s_{i}$ does allow for $h_{i}$ in the tree $T_{h_{i}}$ then there exists no $h_{i}$-replacement of $s_{i}$ which yields player $i$ a higher expected payoff in $T_{h_{i}}$ given the belief $b_{i}\left(h_{i}\right)$ on the other players' strategies $S_{-i}^{T_{h_{i}}}$.

The definition of strong rationalizability is stated next. It is the notion of strong rationalizability of Pearce (1984) and Battigalli (1997) extended to dynamic games with unawareness by Heifetz, Meier, and Schipper (2013). The literature also refers to it as ``extensive-form rationalizability''.

\begin{defin}[Strong rationalizable strategies]\label{rationalizability} For every player $i \in I$, let
\begin{equation*}
{S}_{i}^{0} = S_{i}.
\end{equation*}
For all $k\geq 1$ define recursively, the following sequence of belief systems and strategies of player $i \in I$.
\begin{align*} B_{i}^{1} & = B_{i} \\
S_{i}^{1} & = \left\{s_{i}\in S_{i}: \begin{array}{l} \exists b_{i}\in B_{i}^{1} \ \forall  h_{i}\in H_{i} \ (s_i \mbox{ is rational at } h_i \mbox{ for } b_{i})\end{array}\right\} \\
& \vdots \\
B_{i}^{k} & = \left\{b_{i}\in B_{i}^{k-1}: \begin{array}{l} \forall h_i \in H_i \ \left( \left( S_{-i}^{k-1, T_{h_i}} \cap S_{-i}^{T_{h_i}}(h_i) \neq \emptyset \right) \Longrightarrow \left(b_{i}\left(h_{i}\right)\left(S_{-i}^{k-1, T_{h_{i}}}\right) = 1\right)\right)\end{array} \right\} \\
S_{i}^{k} & = \left\{s_{i}\in S_{i} : \begin{array}{l} \exists b_{i}\in B_{i}^{k} \ \forall  h_{i}\in H_{i} \ (s_i \mbox{ is rational at } h_i \mbox{ for } b_{i}) \end{array} \right\}
\end{align*}

The set of player $i$'s \textbf{strongly rationalizable strategies} is
\begin{equation*}
S_{i}^{\infty} = \bigcap_{k=1}^{\infty } S_{i}^{k}.
\end{equation*}
\end{defin}

Heifetz, Meier, and Schipper (2013) proved that $S_{i}^{k}\subseteq S_{i}^{k-1}$ for every $k > 1$. They also proved that for every finite dynamic game with unawareness, the set of strong rationalizable strategies is non-empty. Battigalli and Siniscalchi (2002) characterized strong rationalizability by rationality and common strong belief in rationality in standard games. This result has been extended to dynamic games with unawareness by Guarino (2020).

Next, we turn to iterated conditional strict dominance. We say that $s_i \in S_i^T$ is \emph{strictly dominated} in a restriction $Y \subseteq S^T$ if $s_i \in Y_i$, $Y_{-i} \neq \emptyset$, and there exists a mixed strategy $\sigma_{i} \in \Delta(Y_i)$ such that $u_i^T(\sigma_i, s_{-i}) > u_i^T(s_i, s_{-i})$ for all $s_{-i} \in Y_{-i}$.

Denote by $\mathbf{S} = \bigcup_{T \in \mathbf{T}} S^T$ and $\mathbf{S}_i = \bigcup_{T \in \mathbf{T}} S^T_i$.

For $T \hookrightarrow T^{\prime}$ and a $T$-partial strategy $s_i \in S_i^T$, we denote the $T^{\prime}$-partial strategy $s_i^{T^{\prime}} \in S_i^{T^{\prime}}$ induced by $s_i$. For $\tilde{s}_i \in S_i^{T^{\prime }}$, define
\begin{equation*}
[\tilde{s}_i] := \bigcup_{T \hookrightarrow T^{\prime }} \{ s_i \in S_i^T :
s_i^{T^{\prime}} = \tilde{s}_i \}.
\end{equation*}
That is, $[\tilde{s}_i]$ is the set of strategies in $\mathbf{S}_i$ which at
information sets $h_i \in H_i^{T^{\prime }}$ prescribe the same actions as
strategy $\tilde{s}_i$. Intuitively, $[\tilde{s}_i]$ is the set of $T$-partial strategies, for any $T \in \mathbf{T}$ with $T \hookrightarrow T'$, that at information sets in the $T'$-partial game are behaviorally indistinguishable from the $T'$-partial strategy $\tilde{s}_i$.

Let $(Y^T)_{T \in \mathbf{T}}$ be a collection of restrictions, one for each $T \in \mathbf{T}$. Define $\mathbf{Y} = \bigcup_{T \in \mathbf{T}} Y^T$. We call this set the extended restriction. Likewise, let $\mathbf{Y}_i = \bigcup_{T \in \mathbf{T}} Y_i^T$, where each $Y_i^T \subseteq S_i^T$ is such that $Y^T = Y_i^T \times Y_{-i}^T$, for some $Y_{-i}^T \subseteq S_{-i}^T$. Intuitively, $\mathbf{Y}_i $ is the projection of $\mathbf{Y} $ to player $i$'s strategy dimension.

\begin{defin}[Conditional Strict Dominance] Given an extended restriction $\mathbf{Y}$, we say that $s_i \in S^T_i$ is \emph{conditionally strictly dominated on} $(\mathcal{X}_i, \mathbf{Y})$ if there exists $T' \in \mathbf{T}$ with $T \hookrightarrow T'$ and $\tilde{s}_i \in S^{T'}_i$ with $s_i \in [\tilde{s}_i]$ such that $\tilde{s}_i$ is strictly dominated
in $X \cap Y^{T'}$ for some normal-form information set $X \in \mathcal{X}_i$, $X \subseteq S^{T'}$.
\end{defin}

This definition implies as a special case that $s_i \in S^T_i$ is conditionally strictly dominated on $(\mathcal{X}_i, \mathbf{Y})$ if there exists a normal-form information set $X \in \mathcal{X}_i$, $X \subseteq S^T$ such that $s_i$ is strictly dominated in $X \cap Y^T$. In particular, when there is just one tree like in standard games without unawareness, the definition reduces to standard conditional dominance. Yet, the domination ``across'' normal forms makes the definition a non-trivial generalization of conditional strict dominance in standard games. When player $i$'s $T'$-partial strategy is strictly dominated conditional on her normal-form information set in the $T'$-partial normal form, then also her strategies with the same components as the $T'$-partial strategy should be eliminated. This is due to the fact that her opponents in $T$-partial games with $T \hookrightarrow T'$ should realize that player $i$ would not play such strategies. That is, a player with a higher awareness level can realize that a strategy is dominated for an opponent with lower awareness level even though that strategy would not be dominated if the opponent were to have the same awareness level as the player. An example is strategy ``$MS$'' for Colin in Section~\ref{Example}. This strategy is not dominated in the upper normal form corresponding to the full game with unawareness. However, its component ``$S$'' is dominated in the $T'$-partial normal form at the bottom of Figure~\ref{modBoSNF}. 

Define an operator on extended restrictions as follows: For any extended restriction $\mathbf{Y}$,
\begin{equation*}
U_{i}(\mathbf{Y}) = \{s_{i} \in \mathbf{Y}_{i}: s_{i}
\mbox{ is not
conditionally strictly dominated on }(\mathcal{X}_{i},\mathbf{Y})\},
\end{equation*}
\begin{equation*}
U(\mathbf{Y}) = \bigcup_{T \in \mathbf{T}} \prod_{i\in I} \left(U_{i}(%
\mathbf{Y}) \cap S_i^T \right),
\end{equation*}
and
\begin{equation*}
U_{-i}(\mathbf{Y}) = \bigcup_{T \in \mathbf{T}} \prod_{j \in I \setminus{%
\{i\}}} \left(U_{j}(\mathbf{Y}) \cap S_j^T \right).
\end{equation*}

The following procedure formalizes iterated elimination of conditionally strictly dominated strategies.

\begin{defin}[Iterated Conditional Strict Dominance] Define recursively,
\begin{align*} U^{0}(\mathbf{S}) & =\mathbf{S},\\
U^{k+1}(\mathbf{S}) & = U(U^{k}(\mathbf{S})) \mbox{ for all } k \geq 0, \\
U^{\infty }(\mathbf{S}) & =\bigcap_{k=0}^{\infty }U^{k}(\mathbf{S}),
\end{align*} and similarly for $U_i^k(\mathbf{S})$ and $U_{-i}^k(\mathbf{S})$. The set $U^{\infty}(\mathbf{S})$ is the maximal reduction under iterated elimination of conditional strictly dominated strategies.
\end{defin}

This procedure generalizes iterated conditional strict dominance by Shimoji and Watson (1998) to games with unawareness.\\

\noindent \textbf{Example (Continued) } We now illustrate the above definitions using the introductory example of Section~\ref{Example}. In Figure~\ref{modBoSNF}, the entire upper normal form is the normal-form information set (marked green) of Rowena (but not Colin) associated with Rowena's information set at the beginning of the $T$-partial game (but not in the $T'$-partial game). We denote this information set by $X_{R}(\emptyset^{T})$. The upper green rectangle in the upper normal form is the normal-form information set of Rowena (but not of Colin) corresponding to her information set after the history $n$ in the $T$-partial game (but not in the $T'$-partial game). We denote it by $X_{R}(n)$. The lower purple rectangle in the upper normal form is the normal-form information set for both Rowena and Colin corresponding to the information sets after history $t$ in the $T$-partial game (but not in the $T'$-partial game). We denote it by $X_{i}(t)$ for $i \in \{C, R\}$. 

Finally, the lower game in normal form is a normal-form information set (marked purple) corresponding to both Colin's and Rowena's information set in the $T'$-partial game. It is also the normal-form information set for Colin corresponding to his information set $\pi_{C}(n)$ in the $T$-partial game. We indicate this with the blue arrow. We denote it by $X_i(\emptyset^{T'}) = X_{C}(n)$.

The definition of $\mathbf{S}_i$ is illustrated by the example $\mathbf{S}_{C} = \{BB, BS, SB, SS, MB, MS, B, S\}$, while the definition of $[\tilde{s}_i]$ can be illustrated by $[``S"] = \{BS, SS, MS, S\}$. These are all the strategies of Colin that prescribe the action ``$S$'' at his information set $\pi_{C}(n)$.

Iterated elimination of conditionally strictly dominated strategies proceeds as follows:
\begin{eqnarray*}
U_{i}^{0}(\mathbf{S}) & = & \mathbf{S}_{i} \mbox{ for } i \in \{R, C\}, \\
U_{R}^{1}(\mathbf{S}) & = & \{nMBB, nMSB, nMMB, nMBS, nMSS, nMMS, \\
&& tBMB, tSMB, tMMB, tBMS, tSMS, tMMS, \\
&& B, S \}, \\
U_{C}^{1}(\mathbf{S}) & = & \{MB,B\}.
\end{eqnarray*}
For instance, strategy $nSBB$ is conditionally strictly dominated by $nMBB$ in the normal-form information set $X_{R}(\emptyset^{T})$ or $X_{R}(n)$. More interestingly, $M S$ is conditionally strictly dominated on $(\mathcal{X}_{C},\mathbf{S})$ because $MS \in [``S"]$ and $S$ is strictly dominated by $B$ in $X_{C}(n)$. This makes sense because a player who is fully aware of the game should realize that when Colin is only aware of the $T'$-partial game, he considers $S$ to be strictly dominated. Thus, Colin will not play strategy $MS$. This example demonstrates that \emph{a strategy in the upper normal form may be eliminated because of strict dominance in the lower normal form.} This is one reason why we chose this game to demonstrate iterated conditional strict dominance in the dynamic game with unawareness. 

Applying the definitions iteratively yields
\begin{eqnarray*}
U_{R}^2(\mathbf{S}) & = & \{nMBB, nMSB, nMMB, B \}, \\
& = & U_{R}^k(\mathbf{S}) \mbox{ for all } k \geq 2, \\
U_{C}^2(\mathbf{S}) & = & U_{C}^2(\mathbf{S}) = \{MB, B\}, \\
& = & U_{C}^k(\mathbf{S}) \mbox{ for all } k \geq 1.
\end{eqnarray*}

Note that $U_{i}^{\infty }\left( \mathbf{S}\right) \cap S_{i} = S_{i}^{\infty }$. That is, the set of strategies remaining after iterated elimination of conditionally strictly dominated strategies coincides with the set of strong rationalizable strategies. Both solution concepts predict that Rowena will play $n$ followed by $M$, while Colin takes action $B$.\hfill $\Box$\\

Our main result of this section is that iterated elimination of conditionally strictly dominated strategies characterizes strong rationalizability in dynamic games with unawareness.

\begin{theo}\label{IECSDS-EFR} For every finite dynamic game with unawareness, $U_{i}^{k}\left( \mathbf{S}\right) \cap S_{i} = S_{i}^{k}$, for all $k \geq 1$. Consequently, $U_{i}^{\infty }\left( \mathbf{S}\right) \cap S_{i} = S_{i}^{\infty }$.
\end{theo}

The proof is contained in the appendix. It proceeds by induction and makes use of Lemma 3 by Pearce (1984). Theorem~\ref{IECSDS-EFR} generalizes Shimoji and Watson (1998) to dynamic games with unawareness.

\section{Prudent Rationalizability and Iterated Conditional Weak Dominance\label{PR}}

Heifetz, Meier, and Schipper (2021) defined prudent rationalizability in order to deal with incautious beliefs in strong rationalizability.

\begin{defin}[Prudent rationalizability]\label{PR_definition} For every player $i \in I$, let
\begin{equation*}
\bar{S}_{i}^{0} = S_{i}.
\end{equation*}
For all $k\geq 1$ define recursively,\footnote{For any probability measure $\mu$ on a finite space $\Omega$, we denote by $supp (\mu) := \{\omega \in \Omega : \mu(\{\omega\}) > 0\}$ the support of $\mu$.}
\begin{align*}
\bar{B}_{i}^{k}\bigskip & = \left\{ b_{i}\in B_{i}: \begin{array}{l} \forall h_i \in H_i \left( \left( \bar{S}_{-i}^{k-1, T_{h_i}} \cap S_{-i}^{T_{h_i}}(h_i) \neq \emptyset \right) \Longrightarrow  \left(supp\left(b_{i}(h_{i})\right) = \bar{S}_{-i}^{k-1,T_{h_{i}}} \cap S_{-i}^{T_{h_i}}(h_i)\right)\right) \end{array}\right\} \\
\bar{S}_{i}^{k} & =  \left\{ s_{i}\in \bar{S}_{i}^{k-1}:
\begin{array}{l} \exists b_{i}\in \bar{B}_{i}^{k} \ \forall h_{i}\in H_{i} \ (s_i \mbox{ is rational at } h_{i} \mbox{ for } b_i)
\end{array}\right\}
\end{align*}

The set of \textbf{prudent rationalizable strategies} of player $i$ is
\begin{equation*}
\bar{S}_{i}^{\infty } = \bigcap_{k=1}^{\infty }\bar{S}_{i}^{k}
\end{equation*}
\end{defin}

Heifetz, Meier, and Schipper (2021) proved that the set of prudent rationalizable strategies is non-empty and discussed various properties and examples. It has been useful in solving disclosure games with and without unawareness (Heifetz, Meier, and Schipper, 2021, Li and Schipper, 2020, 2025), electoral campaigning under unawareness (Schipper and Woo, 2019), and screening and disclosure under unawareness (Francetich and Schipper, 2025). Here we show how to capture this solution concept by iterated elimination of conditional weakly dominated strategies.

We say that $s_i \in S_i^T$ is \emph{weakly dominated} in a restriction $Y \subseteq S^T$ if $s_i \in Y_i$, $Y_{-i} \neq \emptyset$, and there exists a mixed strategy $\sigma_{i} \in \Delta(Y_i)$ such that $u_i^T(\sigma_i, s_{-i}) \geq u_i^T(s_i, s_{-i})$ for all $s_{-i} \in Y_{-i}$ and $u_i^T(\sigma_i, s_{-i}) > u_i^T(s_i, s_{-i})$ for some $s_{-i} \in Y_{-i}$.

\begin{defin}[Conditional Weak Dominance] Given an extended restriction $\mathbf{Y}$, we say that $s_i \in S^T_i$ is \emph{conditionally weakly dominated on} $(\mathcal{X}_i, \mathbf{Y})$ if there exists $T' \in \mathbf{T}$ with $T \hookrightarrow T'$ and $\tilde{s}_i \in S^{T'}_i$ with $s_i \in [\tilde{s}_i]$ such that $\tilde{s}_i$ is weakly dominated in $X \cap Y^{T^{\prime }}$ for some normal-form information set $X \in \mathcal{X}_i$, $X \subseteq S^{T^{\prime }}$.
\end{defin}

Note that this definition implies as a special case that $s_i \in S^T_i$ is conditionally weakly dominated on $(\mathcal{X}_i, \mathbf{Y})$ if there exists a normal-form information set $X \in \mathcal{X}_i$, $X \subseteq S^T$ such that $s_i$ is weakly dominated in $X \cap Y^T$. Yet, the weak domination ``across'' normal forms makes this definition a non-trivial generalization of conditional weak dominance. Again, the idea is that a player who perceives his opponent to be aware of less strategies than himself may realize that the opponent considers some of her partial strategies as being weakly dominated. Consequently, the player knows that his opponent will not play any strategy that has the weakly dominated partial strategies as components.

We define an operator on extended restrictions as follows: For any extended restriction $\mathbf{Y}$,
\begin{align*}
W_{i}(\mathbf{Y}) & = \{s_{i} \in \mathbf{Y}_{i}: s_{i}
\mbox{ is not conditionally weakly dominated on }(\mathcal{X}_{i},\mathbf{Y})\},
\end{align*}
\begin{align*}
W(\mathbf{Y}) & =  \bigcup_{T \in \mathbf{T}} \prod_{i\in I} \left(W_{i}(\mathbf{Y}) \cap S_i^T \right),
\end{align*}
and
\begin{align*}
W_{-i}(\mathbf{Y}) & =  \bigcup_{T \in \mathbf{T}} \prod_{j \in I \setminus{\{i\}}} \left(W_{j}(\mathbf{Y}) \cap S_j^T \right).
\end{align*}

\begin{defin}[Iterated Elimination of Conditionally Weakly Dominated Strategies] Define recursively,
\begin{align*}
W^{0}(\mathbf{S}) & = \mathbf{S},\\
W^{k+1}(\mathbf{S}) & = W(W^{k}(\mathbf{S})) \mbox{ for all } k \geq 0,\\
W^{\infty}(\mathbf{S}) & = \bigcap_{k=0}^{\infty} W^{k}(\mathbf{S}),
\end{align*} and similarly for $W_i^k(\mathbf{S})$ and $W_{-i}^k(\mathbf{S})$. The set $W^{\infty}(\mathbf{S})$ is the maximal reduction under iterated elimination of conditionally weakly dominated strategies.
\end{defin}

This extends the notion of iterated elimination of conditionally weakly dominated strategies to games with unawareness. We characterize prudent rationalizability by iterated elimination of conditionally weakly dominated strategies.

\begin{theo}\label{IECWDS-PR} For every finite dynamic game with unawareness, $W_{i}^{k}\left( \mathbf{S}\right) \cap S_{i} = \bar{S}_{i}^{k}$, for all $k \geq 1$. Consequently, $W_{i}^{\infty }\left( \mathbf{S}\right) \cap S_{i} = \bar{S}_{i}^{\infty }$.
\end{theo}

\noindent \textbf{Proof.} The proof is analogous to Theorem~\ref{IECSDS-EFR} in the appendix. Instead of using Lemma 3 of Pearce (1984), we now use Lemma 4 of Pearce (1984). Moreover, instead of nestedness of beliefs across rounds, we directly use the nestedness of prudent rationalizable strategies across rounds. \hfill $\Box$

\section{Iterated Admissibility\label{IA}}

For standard games in extensive form with perfect recall and without unawareness, Brandenburger and Friedenberg (2011, Proposition 3.1) showed that iterated elimination of conditionally weakly dominated strategies coincides with iterated admissibility at each iteration in the associated normal form. Before we can extend the result to games with unawareness, we first need a notion of iterated admissibility for games with unawareness. How to define iterated admissibility in games with unawareness in the generalized normal form?\\

\noindent \textbf{Example (Continued) } To motivate our notion of iterated admissibility for games with unawareness, we consider again the example in Section~\ref{Example}. A straightforward but naive approach could be to apply iterated admissibility to each of the two normal forms separately. At the first level, the set of admissible strategies coincides with the set of strategies that are not conditionally strictly dominated except for $MS$ of Colin. (Recall that under conditional strict dominance, we were able to delete strategy $MS$ because $S$ was strictly dominated in the lower game $T'$.) Notice that at the second level, strategy $MB$ now becomes weakly dominated by $MS$. But $MB$ is the only strongly rationalizable strategy and prudent rationalizable strategy of Colin in the $T$-partial game. What is wrong with this naive approach is that it does not eliminate a strategy in the $T$-partial game when it is weakly dominated by another strategy in a lower game. Strategies $MB$ and $MS$ differ in the second component only, the action of Colin in the lower game $T'$. It is in this lower game $T'$ that $S$ is dominated by $B$, and hence any strategy prescribing $S$ at the lower game $T'$ should be eliminated. This motivates us to define iterated admissibility as a procedure that \emph{conditions at least on normal forms} instead of any normal-form information set. But not any normal form will do. We also need to ensure that for each information set, we condition on the ``correct'' normal form, namely the normal form that represents the player's awareness at this information set. This highlights an additional role of information sets in dynamic games with unawareness. A player's information set embodies not only information in the standard sense but also the player's awareness that is given by the tree in which this information set is located. \hfill $\Box$\\

More formally, let $\mathcal{S}_i = \{ S^{T_{h_i}} : h_i \in H_i \}$. This is the collection of all strategy profiles and $T$-partial strategy profiles for any tree $T$ for which player $i$ has an information set. We can understand it as the set of normal forms of player $i$.

\begin{defin}[NF-conditional Weak Dominance] Given an extended restriction $\mathbf{Y}$, we say that $s_i \in S^T_i$ is \emph{NF-conditionally weakly dominated on} $(\mathcal{S}_i, \mathbf{Y})$ if there exists $T' \in \mathbf{T}$ such that $T \hookrightarrow T'$, $S^{T'} \in \mathcal{S}_i$, and $\tilde{s}_i \in S^{T'}_i$ with $s_i \in [\tilde{s}_i]$, such that $\tilde{s}_i$ is weakly dominated in $S^{T'} \cap Y^{T^{\prime }}$.
\end{defin}

Note that this definition implies as a special case that $s_i \in S^T_i$ is NF-conditionally weakly dominated on $(\mathcal{S}_i, \mathbf{Y})$, if $s_i$ is weakly dominated in $S^T \cap Y^T = Y^T$. Again, the weak domination ``across'' normal forms makes this definition a non-trivial generalization of weak dominance.

Define an operator on extended restrictions as follows: For any extended restriction $\mathbf{Y}$,
\begin{equation*}
\widetilde{W}_{i}(\mathbf{Y}) = \{s_{i} \in \mathbf{Y}_{i}: s_{i}
\mbox{ is not NF-conditionally weakly dominated on }(\mathcal{S}_{i},\mathbf{Y})\},
\end{equation*}
\begin{equation*}
\widetilde{W}(\mathbf{Y}) = \bigcup_{T \in \mathbf{T}} \prod_{i\in I} \left(\widetilde{W}_{i}(\mathbf{Y}) \cap S_i^T \right),
\end{equation*}
and
\begin{equation*}
\widetilde{W}_{-i}(\mathbf{Y}) = \bigcup_{T \in \mathbf{T}} \prod_{j \in I \setminus{\{i\}}} \left(\widetilde{W}_{j}(\mathbf{Y}) \cap S_j^T \right).
\end{equation*}
Compared to operator $W$ defined earlier that considered strategies that are not conditionally weakly dominated, operator $\widetilde{W}$ considers strategies that are not $NF$-conditionally weakly dominated. This means that with $\widetilde{W}$ we condition only on the normal form but not on finer subsets of strategy profiles.

\begin{defin}[Iterated Admissibility] Define recursively,
\begin{align*} \widetilde{W}^{0}(\mathbf{S}) & = \mathbf{S},\\
\widetilde{W}^{k+1}(\mathbf{S}) & =\widetilde{W}(\widetilde{W}^{k}(\mathbf{S})) \mbox{ for all } k\geq 0,\\
\widetilde{W}^{\infty }(\mathbf{S}) & = \bigcap_{k=0}^{\infty }\widetilde{W}^{k}(\mathbf{S}),
\end{align*} and similarly for $\widetilde{W}_i^k(\mathbf{S})$ and $\widetilde{W}_{-i}^k(\mathbf{S})$. The set $\widetilde{W}^{\infty }(\mathbf{S})$ is the maximal reduction under iterated admissibility. We call any strategy of player $i$ in $\widetilde{W}_{i}(\mathbf{Y})$ admissible on $(\mathcal{S}_{i},\mathbf{Y})$, and any strategy of player $i$ in $\widetilde{W}_i^{\infty }(\mathbf{S})$ as iterated admissible.
\end{defin}

This procedure generalizes iterated admissibility to games with unawareness. While the definition boils down to admissibility (resp., iterated admissibility) when restricted to standard games without unawareness, the terminology, while correct, may be somewhat misleading under unawareness. In games with unawareness, iterated admissibility is conceptually closer to iterated conditional weak dominance because it makes explicit use of the awareness also embodied in information sets. In dynamic games with unawareness, the information set contains not only information in the standard sense but also awareness given by the tree on which it is located. This awareness determines the relevant normal form that is conditioned on in the definition of iterated admissibility for games with unawareness. That is, while iterated admissibility in standard games does not rely on any structure in the extensive form, iterated admissibility in dynamic games must rely on the awareness captured by information sets in the extensive form. Iterated admissibility as defined here allows us to neglect information but not awareness embodied in the extensive form.

We show that prudent rationalizability is captured also by iterated admissibility in dynamic games with unawareness.

\begin{theo}\label{IA-PR} For every finite dynamic game with unawareness, $\widetilde{W}_{i}^{k}\left( \mathbf{S}\right) = \bar{S}_i^k$, for all $k\geq 1$. Consequently, $\widetilde{W}_{i}^{\infty }\left( \mathbf{S}\right) = \bar{S}_i^{\infty}$.
\end{theo}

The proof is contained in the appendix. There, we first define a version of prudent rationalizability that weakens dynamic properties of belief systems. We then show inductively the equivalence between that version of prudent rationalizability and iterated admissibility, again using Lemma 4 in Pearce (1984). Finally, we show inductively the equivalence between that weakened version of prudent rationalizability and prudent rationalizability of Definition~\ref{PR_definition}. 

As a corollary, iterated admissibility is equivalent to iterated conditional weak dominance at every iteration in dynamic games with unawareness. This result generalizes a result for standard games by Brandenburger and Friedenberg (2011, Proposition 3.1) to dynamic games with unawareness.

\begin{cor}\label{IECWDS-IA} For every finite dynamic game with unawareness, $\widetilde{W}_{i}^{k}\left( \mathbf{S}\right) = W_{i}^{k}\left( \mathbf{S}\right)$, for all $k\geq 1$. Consequently, $\widetilde{W}_{i}^{\infty }\left( \mathbf{S}\right) = W_{i}^{\infty }\left( \mathbf{S}\right)$.
\end{cor}

\appendix

\section{Properties of Information Sets\label{properties}}

To make the paper self-contained, we list properties of information sets satisfied by dynamic games with unawareness; see Heifetz, Meier, and Schipper (2013) and Schipper (2019, 2021) for further discussions and illustrations. Some of these properties are explicitly used in the proofs of the results. 

For each decision node $n \in N$ and active player $i \in I_{n}$ there is a nonempty information set $\pi_{i}\left(n\right)$ satisfying the following properties:

\begin{itemize}
\item[I0] Confinement: $\pi _{i}\left( n\right) \subseteq T$ for some tree $T
$.

\item[I1] No-delusion given the awareness level: If $\pi _{i}(n)\subseteq
T_{n}$, then $n \in \pi_{i}(n)$.

\item[I2] Introspection: If $n^{\prime} \in \pi_{i}\left( n\right)$, then $%
\pi _{i}\left( n^{\prime }\right) = \pi _{i}\left( n\right)$.

\item[I3] No divining of currently unimaginable paths, no expectation to
forget currently conceivable paths: If $n^{\prime }\in \pi_{i}\left(
n\right) \subseteq T^{\prime }$ (where $T^{\prime }\in \mathbf{T}$ is a
tree) and there is a path $n^{\prime },\dots , n^{\prime \prime }\in
T^{\prime }$ such that $i\in I_{n^{\prime }}\cap I_{n^{\prime \prime }}$,
then $\pi _{i}\left( n^{\prime \prime }\right) \subseteq T^{\prime }$.

\item[I4] No imaginary actions: If $n^{\prime }\in \pi _{i}\left( n\right)$,
then $A_{n^{\prime }}^{i}\subseteq A_{n}^{i}$.

\item[I5] Distinct action names in disjoint information sets: For a subtree $T$, if $n, n^{\prime } \in T$ and $A_{n}^{i}=A_{n^{\prime }}^{i}$, then $\pi_{i}\left( n^{\prime}\right) =\pi_{i}\left( n\right)$.

\item[I6] Perfect recall: Suppose that player $i$ is active at two distinct
nodes $n_{1}$ and $n_{k}$, and there is a path $n_{1}, n_{2}, ..., n_{k}$
such that at $n_{1}$ player $i$ takes the action $a_{i}$. If $n^{\prime} \in
\pi _{i}\left( n_{k}\right)$, then there exists a node $n_{1}^{\prime}\neq
n^{\prime }$ and a path $n_{1}^{\prime}, n_{2}^{\prime }, ...,
n_{\ell}^{\prime } = n^{\prime }$ such that $\pi_{i}\left( n_{1}^{\prime
}\right) = \pi_{i} \left( n_{1}\right) $ and at $n_{1}^{\prime }$ player $i$
takes the action $a_{i}$.
\end{itemize}

\section{Proofs}

\subsection*{Proof of Theorem~\ref{IECSDS-EFR}}

A \emph{general belief system} of player $i \in I$,
\begin{equation*}
\widehat{b}_{i}=(\widehat{b}_{i}(h_{i}))_{h_{i}\in H_{i}}\in \prod_{h_{i}\in
H_{i}}\Delta (S_{-i}^{T_{h_{i}}})
\end{equation*}
is a profile of beliefs -- a belief $\widehat{b}_{i}(h_{i})\in \Delta(S_{-i}^{T_{h_{i}}})$ about the other players' strategies in the $T_{h_{i}}$-partial game, for each information set $h_{i}\in H_{i}$,
such that $\widehat{b}_{i}(h_{i})$ assigns probability $1$ to the set of strategy profiles of the other players that allow for $h_{i}$. The difference between a belief system and a general belief system is that for the latter \emph{we do not impose conditioning}.

For $i \in I$, let  $\widehat{S}_{i}^{0} = {S}_{i}$, and for $k \geq 1$, let $\widehat{B}_{i}^{k}$ and $\widehat{S}_{i}^{k}$ be defined recursively like $B_{i}^{k}$ and $S_{i}^{k}$ in Definition~\ref{rationalizability}, respectively, the only change being that belief systems are replaced by generalized belief systems.

\begin{lem}\label{DS} $U_{i}^{k}(\mathbf{S}) \cap S_i =\widehat{S}_{i}^{k}$
for all $k\geq 0$. Consequently, $U_{i}^{\infty }(\mathbf{S}) \cap S_i =\widehat{S}%
_{i}^{\infty} $.
\end{lem}

\noindent \textbf{Proof of the Lemma~\ref{DS}. } We proceed by
induction. The case $k = 0$ is straightforward since $U_i^0(\mathbf{S}) \cap S_i = S_i = \widehat{S}_i^0$ for all $i \in I$.

Suppose now that we have shown $U_{i}^{\ell}(\mathbf{S}) \cap S_i =\widehat{S}_{i}^{\ell}$ for all $i \in I$ and $\ell \leq k$. We want to show that $U_{i}^{k+1}(\mathbf{S}) \cap S_i =\widehat{S}_{i}^{k+1}$ for all $i \in I$.

\noindent ``$\subseteq$'': First we show, if $s_i \in U_i^{k+1}(\mathbf{S}) \cap S_i$, then $s_i \in \widehat{S}_i^{k + 1}$.

We have $s_i \in U_i^{k+1}(\mathbf{S}) \cap S_i$ if and only if $s_i \in U_i^{k}(\mathbf{S}) \cap S_i$ and $s_i$ is not conditionally strictly dominated on $(\mathcal{X}_i, U^k(\mathbf{S}))$. Strategy $s_i \in U_i^{k}(\mathbf{S}) \cap S_i$ is not conditionally strictly dominated on $(\mathcal{X}_i, U^k(\mathbf{S}))$ if for all $T^{\prime }\in \mathbf{T}$ with $T_1 \hookrightarrow T^{\prime }$ and all $\tilde{s}_i \in S_i^{T^{\prime }}$ such that $s_i \in [\tilde{s}_i]$, we have that there does not exist a normal-form information set $X \in \mathcal{X}_i$ with $X \subseteq S^{T^{\prime }}$ such that $\tilde{s}_i$ is strictly dominated in $X \cap U^k(\mathbf{S}) \cap S^{T^{\prime }}$. For any information set $h_i \in H_i$, if $\tilde{s}_i \in S_i^{T_{h_i}}$ is not strictly dominated in $S^{T_{h_i}}(h_i) \cap U^k(\mathbf{S}) \cap S^{T^{\prime }}$, then either

\begin{itemize}
\item[(i)] $\tilde{s}_i$ does not allow for $h_i$, or

\item[(ii)] $\tilde{s}_i$ does allow for $h_i$ but $S_{-i}^{T_{h_i}}(h_i) \cap U^k_{-i}(\mathbf{S}) = \emptyset$; or 

\item[(iii)] $\tilde{s}_i$ does allow for $h_i$ and $S_{-i}^{T_{h_i}}(h_i) \cap U^k_{-i}(\mathbf{S}) \neq \emptyset$. In this case, 
by Lemma 3 in Pearce (1984) there exists a belief, let's denote it by $d_i(h_i)$, with $d_i(h_i)\left(S_{-i}^{T_{h_i}}(h_i) \cap U^k_{-i}(\mathbf{S})\right) = 1$ and for which $\tilde{s}_i$ is rational at $h_i$. Since by the induction hypothesis $U^k(\mathbf{S}) \cap S = \widehat{S}^k$, we have $d_i(h_i)\left(\widehat{S}_{-i}^{k, T_{h_i}}(h_i)\right) = 1$.
\end{itemize}

We need to construct a generalized belief system in $\widehat{b}_i \in \widehat{B}_i^{k+1}$ such that $\tilde{s}_i$ is rational at every $h_i \in H_i$ for $\widehat{b}_i(h_i)$. By the induction hypothesis, $\tilde{s}_i \in U_i^{k}(\mathbf{S}) \cap S_i = \widehat{S}_i^k$. Thus, there exists a generalized belief system $\widehat{b}'_i = (\widehat{b}'_i(h_i))_{h_i \in H_i} \in \widehat{B}_i^{k}$ such that for all $h_i \in H_i$, $\tilde{s}_i$ is rational at $h_i$ for $\widehat{b}'_i(h_i)$. For any $h_i \in H_i$ satisfying case (iii), replace $\widehat{b}'_i(h_i)$ by $d_i(h_i)$ as defined in (iii). For all other  $h_i \in H_i$, let $\widehat{b}_i(h_i) := \widehat{b}_i'(h_i)$. This yields the generalized belief system $\widehat{b}_i = (\widehat{b}_i(h_i))_{h_i \in H_i} \in \widehat{B}_i^{k + 1}$. 

Observe that for any $h_i \in H_i$ satisfying case (iii), $\tilde{s}_i$ is rational at $h_i$ for $\widehat{b}_i(h_i)$ by the arguments made under case (iii). For any $h_i \in H_i$  satisfying case (ii), we also have $\tilde{s}_i$ is rational at $h_i$ for $\widehat{b}_i(h_i)$ because in this case $\widehat{b}_i(h_i) = \widehat{b}'_i(h_i)$, $\widehat{b}'_i \in \widehat{B}_i^k$, and $\tilde{s}_i$ is rational at $h_i$ for $\widehat{b}'_i(h_i)$ by assumption. For any $h_i \in H_i$ satisfying case (i), $\tilde{s}_i$ is trivially rational at $h_i$ for $\widehat{b}_i(h_i)$. 

Note that by the definitions of $[\tilde{s}_i]$ and ``allow for'', if $\tilde{s}_i$ allows for $h_i$ in the tree $T_{h_i}$ and $s_i \in [\tilde{s}_i]$, then $s_i$ allows for $h_i$ in the tree $T_{h_i}$. Hence, if $\tilde{s}_i \in S_i^{T_{h_i}}$ is rational at $h_i$ given $\widehat{b}_i(h_i)$, then $s_i \in [\tilde{s}_i]$ is rational at $h_i$ given $\widehat{b}_i(h_i)$.

We conclude $s_i \in \widehat{S}_i^{k+1}$.\newline

\noindent ``$\supseteq$'': We show that, if $s_i \in \widehat{S}_i^{k + 1}$, then $s_i \in U_i^{k+1}(\mathbf{S}) \cap S_i$.

If $s_i \in \widehat{S}_i^{k+1}$, then there exists a generalized
belief system $\widehat{b}_i \in \widehat{B}_i^{k+1}$ such that for all $h_i \in
H_i$ the strategy $s_i$ is rational given $\widehat{b}_i(h_i)$. This implies that for any $h_i \in H_i$ with $\widehat{S}_{-i}^{k, T_{h_i}} \cap S_{-i}^{T_{h_i}}(h_i) \neq \emptyset$ we have $\widehat{b}_i(h_i)\left(\widehat{S}_{-i}^{k, T_{h_i}}(h_i)\right) = 1$. By the induction hypothesis, $\widehat{S}^{k, T_{h_i}}_{-i}(h_i) = U_{-i}^k(\mathbf{S}) \cap S_{-i}^{T_{h_i}}(h_i)$. Hence $\widehat{b}_i(h_i)\left(
U_{-i}^k(\mathbf{S}) \cap S_{-i}^{T_{h_i}}(h_i)\right) = 1$.

If $s_i \in [\tilde{s}_i]$ with $\tilde{s}_i \in S_i^{T_{h_i}}$ and $s_i$ allows for $h_i$ in the tree $T_{h_i}$, then $\tilde{s}_i$ allows for $h_i$ in the tree $T_{h_i}$. Hence, if $s_i \in [\tilde{s}_i]$ with $\tilde{s}_i \in S_i^{T_{h_i}}$ is rational at $h_i$ given $\widehat{b}_i(h_i)$, then $\tilde{s}_i$ is rational at $h_i$ given $\widehat{b}_i(h_i)$. Thus, if $s_i$ is rational at $h_i$ given $\widehat{b}_i(h_i)$,
then $\tilde{s}_i \in S_i^{T_{h_i}}$ with $s_i \in [\tilde{s}_i]$ is not strictly dominated in $U_{-i}^k(\mathbf{S}) \cap S_{-i}^{T_{h_i}}(h_i)$
either because $s_i$ does not allow for $h_i$, or because $U_{-i}^k(\mathbf{S}) \cap S_{-i}^{T_{h_i}}(h_i) = \emptyset$, or because of Lemma 3 in Pearce (1984). It then follows that if there exists a generalized belief system $\widehat{b}_i \in \widehat{B}_i^{k+1}$ such that for all $h_i \in H_i$ the strategy $s_i$ is rational given $\widehat{b}_i(h_i)$, then $s_i$ is not conditionally strictly
dominated on $(\mathcal{X}_i, U^k(\mathbf{S}))$. Hence $s_i \in U_i^{k+1}(\mathbf{S}) \cap S_i$.\hfill $\Box$

\begin{lem}\label{inconsistent} $\widehat{S}_{i}^{k} = S_{i}^{k}$ for all $k \geq 1$. Consequently, $\widehat{S}_{i}^{\infty} = S_{i}^{\infty}$.
\end{lem}

\noindent \textbf{Proof of the Lemma~\ref{inconsistent}. } We have $S_{i}^{k}\subseteq \widehat{S}_{i}^{k}$ for all $k\geq 1$ since, if $s_{i}$ is rational at each information set $h_{i}\in H_{i}$ given the belief system $b_{i}\in B_{i}$, then there exists a generalized belief system $\widehat{b}_{i}\in \widehat{B}_{i}^{k}$, namely $\widehat{b}_{i} = b_{i}$, such that $s_{i}$ is rational at each information set $h_{i}\in H_{i}$ given $\widehat{b}_{i}$.

We need to show the reverse inclusion, $\widehat{S}_{i}^{k}\subseteq S_{i}^{k}$ for all $k \geq 1$. The first step is to show how to construct a (consistent) belief system from a generalized belief system. Let $s_{i}$ be rational given $\widehat{b}_{i}\in \widehat{B}_{i}^{1}$, i.e., $s_{i}\in \widehat{S}_{i}^{1}$. Consider an information set $h_{i}^{0}\in H_{i}$ such that in $T_{h_{i}^0}$ there does not exist an information set $h_{i}$ that precedes $h_{i}^{0}$. Define $b_{i}(h_{i}^{0}) := \widehat{b}_{i}(h_{i}^{0})$.

Assume, inductively, that we have already defined $b_{i}$ for a subset of information sets $H_{i}^{\prime }\subseteq H_{i}$ such that for each $h_{i}^{\prime }\in H_{i}^{\prime }$ all the predecessors of $h_{i}^{\prime }$ are also in $H_{i}^{\prime }.$ For each successor
information set $h_{i}^{\prime \prime }$ of each information set $h_{i}^{\prime }\in H_{i}^{\prime }$ such that $h_{i}^{\prime \prime }\notin H_{i}^{\prime }$ define $b_{i}\left( h_{i}^{\prime \prime }\right) $ as follows:
\begin{itemize}
\item If $b_{i}(h_{i}')$ assigns strict positive probability to a strategy profile of other players, $s_{-i}^{T_{h_i'}}$, that allows for $h_{i}''$, then define $b_{i}\left( h_{i}''\right)$ by using conditioning, i.e., if $s_{-i}^{T_{h_i'}} \in S_{-i}(h_i^{\prime \prime })$
\begin{equation*} b_{i}\left( h_{i}^{\prime \prime }\right) \left(s_{-i}^{T_{h_{i}^{\prime }}}\right) =
\frac{b_{i}\left( h_{i}^{\prime }\right) \left(s_{-i}^{T_{h_{i}^{\prime }}}\right)}{\sum_{\tilde{s}_{-i}^{T_{h_{i}^{\prime }}}\in S_{-i}(h_{i}^{\prime \prime
})}b_{i}(h_{i}^{\prime })\left(\tilde{s}_{-i}^{T_{h_{i}^{\prime }}}\right)}.
\end{equation*}

\item Otherwise let $b_{i}(h_{i}^{\prime \prime })\equiv \widehat{b}_{i}(h_{i}^{\prime \prime })$.
\end{itemize}

Since there are finitely many information sets in $H_{i}$, this
recursive definition concludes in a finite number of steps.

Next, assuming that $s_{i}$ is rational at each information set
$h_{i}\in H_{i}$ \ with the generalized belief system $\widehat{b}_{i},$ we
will show that $s_{i}$ is also rational at each information set $h_{i}\in
H_{i}$ according to the belief system $b_{i}$.

Consider again $h_{i}^{0}\in H_{i}$ with no predecessors in $T_{h_{i}^{0}}$. Since $b_{i}(h_{i}^{0})=\widehat{b}_{i}(h_{i}^{0})$ and $s_{i}$ is rational at $h_{i}^{0}$ given $\widehat{b}_{i}\left( h_{i}^{0}\right) $, $s_{i}$ is also rational at $h_{i}^{0}$ given $b_{i}\left( h_{i}^{0}\right)$.

Assume, inductively, that we have already shown the claim for a subset of information sets $H_{i}^{\prime }\subseteq H_{i}$ such that for each $h_{i}^{\prime }\in H_{i}^{\prime }$ all the predecessors of $h_{i}^{\prime }$ are also in $H_{i}^{\prime }$. Consider a successor information set $h_{i}^{\prime \prime }$ of an information set $h_{i}^{\prime }\in H_{i}^{\prime }$ such that $h_{i}^{\prime \prime }\notin H_{i}^{\prime }.$ \ Notice that each $h_{i}^{\prime \prime }$-replacement is also an $h_{i}^{\prime }$-replacement. Therefore,

\begin{itemize}
\item If $b_{i}(h_{i}')$ assigns strict positive probability to a strategy profile of other players, $s_{-i}^{T_{h_i'}}$, that allows for $h_{i}''$, then by above construction $b_{i}\left( h_{i}''\right)$ is derived from $b_{i}(h_{i}')$ by conditioning, and hence any $h_{i}''$-replacement that is improving player $i$'s expected payoff according to $b_{i}\left( h_{i}''\right)$ would also improve
player $i$'s payoff with $b_{i}(h_{i}')$, contradicting the induction hypothesis. Hence $s_{i}$ is rational at $h_{i}''$ given $b_{i}\left( h_{i}''\right)$.

\item Otherwise, by above construction, $b_{i}\left( h_{i}''\right) = \widehat{b}_{i}(h_{i}'')$. Hence, $s_{i}$ is rational at $h_{i}''$ also for $b_{i}\left( h_{i}''\right)$.
\end{itemize}

Applying the same argument inductively yields $\widehat{S}_{i}^{k} = S_{i}^{k}$ for all $k \geq 1$. This concludes the proof of the lemma. \hfill $\Box$ \newline

Lemmata~\ref{DS} and~\ref{inconsistent} together yield $U_{i}^{k}(\mathbf{S})\cap S_{i} = S_{i}^{k}$ for all $k \geq 1$. Since it applies for all $k \geq 1$ and $i \in I$, this completes the proof of the theorem.\hfill $\Box$

\subsection*{Proof of Theorem~\ref{IA-PR}}

The proof proceeds in several steps. First, we show that iterated admissibility is characterized by a variant of prudent rationalizability.

A \emph{relaxed belief system} of player $i \in I$,
\begin{equation*}
\ddot{b}_{i}=\left( \ddot{b}_{i}\left( h_{i}\right) \right) _{h_{i}\in H_{i}}\in
\prod_{h_{i}\in H_{i}}\Delta \left( S_{-i}^{T_{h_{i}}}\right)
\end{equation*}
is a profile of beliefs, a belief $\ddot{b}_{i}\left( h_{i}\right) \in \Delta
\left( S_{-i}^{T_{h_{i}}}\right)$ about the other players' strategies in
the $T_{h_{i}}$-partial game for each information set $h_{i}\in H_{i}$. Compared to general belief systems, for relaxed belief systems we do \emph{not} require that for any $h_i \in H_i$, belief $\ddot{b}_{i}\left( h_{i}\right)$ assigns probability 1 to the set of strategy profiles of the other players in the $T_{h_i}$-partial game that allow for $h_{i}$. We also do \emph{not} require that, if $h_{i}\rightsquigarrow h_{i}'$, then $\ddot{b}_i(h_i')$ is derived from $\ddot{b}_i(h_i)$ by conditioning whenever possible.  

Denote by $\ddot{B}_{i}$ the set of player $i$'s relaxed belief systems.

\begin{defin}[Relaxed prudent rationalizability] Let
\begin{align*}
\ddot{S}_{i}^{0} & = S_{i}
\end{align*}
For all $k \geq 1$, define recursively,
\begin{align*}
\ddot{B}_{i}^{k} & = \left\{ \ddot{b}_{i} \in \ddot{B}_{i}:
\begin{array}{l} \forall h_{i} \in H_i \ \left( \left( \ddot{S}_{-i}^{k-1, T_{h_i}} \cap S_{-i}^{T_{h_i}}(h_i) \neq \emptyset\right) \Longrightarrow \left( supp (\ddot{b}_{i}(h_{i})) =  \ddot{S}_{-i}^{k-1,T_{h_{i}}} \right) \right)
\end{array}\right\} \\
\ddot{S}_{i}^{k} & = \left\{ s_{i}\in \ddot{S}_{i}^{k-1}:
\begin{array}{l}
\exists \ddot{b}_{i} \in \ddot{B}_{i}^{k} \ \forall h_{i}\in H_{i} \  (s_i \mbox{ is rational at } h_i \mbox{ for } \ddot{b}_i)
\end{array}\right\}
\end{align*}

The set of relaxed prudent rationalizable strategies of player $i$ is
\begin{align*}
\ddot{S}_{i}^{\infty} & = \bigcap_{k=1}^{\infty }\ddot{S}_{i}^{k}
\end{align*}
\end{defin}

\begin{prop}\label{IA-RPR} For every finite dynamic game with unawareness, $\widetilde{W}_{i}^{k}\left( \mathbf{S}\right) \cap S_{i} = \ddot{S}_{i}^{k}$, for all $k\geq 1$. Consequently, $\widetilde{W}_{i}^{\infty }\left( \mathbf{S}\right) \cap S_{i} = \ddot{S}_{i}^{\infty}$.
\end{prop}

\noindent \textbf{Proof of Proposition~\ref{IA-RPR}:} We proceed by induction. The case $k = 0$ is straightforward since $\widetilde{W}_i^0(\mathbf{S}) \cap S_i = S_i = \ddot{S}_i^0$ for all $i \in I$.

Suppose now that we have shown $\widetilde{W}_{i}^{k}(\mathbf{S}) \cap S_i =\ddot{S}_{i}^{k}$ for all $i \in I$. We want to show that $\widetilde{W}_{i}^{k+1}(\mathbf{S}) \cap S_i =\ddot{S}_{i}^{k+1}$ for all $i \in I$.

\noindent ``$\subseteq$'': First, we show that, if $s_i \in \widetilde{W}_i^{k+1}(\mathbf{S}) \cap S_i$, then $s_i \in \ddot{S}_i^{k + 1}$.

We have $s_i \in \widetilde{W}_i^{k+1}(\mathbf{S}) \cap S_i$ if $s_i \in S_i$ is not NF-conditionally weakly dominated on $(\mathcal{S}_i, \widetilde{W}^k(\mathbf{S}))$. Strategy $s_i \in S_i$ is not NF-conditionally weakly dominated on $(\mathcal{S}_i, \widetilde{W}^k(\mathbf{S}))$ if for all $T^{\prime }\in \mathbf{T}$ with $T_1 \hookrightarrow T^{\prime }$ and all $\tilde{s}_i \in S_i^{T^{\prime }}$ such that $s_i \in [\tilde{s}_i]$, we have that $\tilde{s}_i$ is not weakly dominated in $S_i^{T^{\prime }} \cap \widetilde{W}^k(\mathbf{S})$.

For any information set $h_i \in H_i$, if $\tilde{s}_i \in S_i^{T_{h_i}}$ is not weakly dominated in $S^{T_{h_i}} \cap \widetilde{W}^k(\mathbf{S})$, then either $\tilde{s}_i$ does not allow for $h_i$, or $S_{-i}^{T_{h_i}}(h_i) \cap \widetilde{W}^k_{-i}(\mathbf{S}) = \emptyset$, or by Lemma 4 in Pearce (1984) there exists a belief $d_i(h_i)$ with $supp(d_i(h_i)) = S_{-i}^{T_{h_i}} \cap \widetilde{W}^k_{-i}(\mathbf{S})$ such that $\tilde{s}_i$ is rational at $h_i$ for $d_i(h_i)$. Since by the induction hypothesis $\widetilde{W}^k(\mathbf{S}) \cap S = \ddot{S}^k$, we have in the latter case that $supp(d_i(h_i)) = \ddot{S}_{-i}^{k, T_{h_i}}$.

Consider a relaxed belief system $\ddot{b}'_i = (\ddot{b}'_i(h_i))_{h_i \in H_i} \in \ddot{B}_i$ and define a new profile of beliefs $\ddot{b}_i = (\ddot{b}_i(h_i))_{h_i \in H_i}$ by $\ddot{b}_i(h_i) = d_i(h_i)$ if $h_i \in H_i$ is such that $\tilde{s}_i$ reaches $h_i$ and $S_{-i}^{T_{h_i}}(h_i) \cap \widetilde{W}^k_{-i}(\mathbf{S}) \neq \emptyset$. Otherwise, let $\ddot{b}_i(h_i) = \ddot{b}'_i(h_i)$. Note that $\ddot{b}_i$ is a relaxed belief system in $\ddot{B}_i^{k+1}$.

By definition of $[\tilde{s}_i]$, if $\tilde{s}_i \in S_i^{T_{h_i}}$ is rational at $h_i$ for $\ddot{b}_i(h_i)$, then $s_i \in [\tilde{s}_i]$ is rational at $h_i$ for $\ddot{b}_i(h_i)$. Finally, we need to check that $s_i \in \ddot{S}_i^{k}$. Note that $s_i \in \widetilde{W}_i^{k+1}(\mathbf{S}) \cap S_i \subseteq \widetilde{W}_i^{k}(\mathbf{S}) \cap S_i$. By the induction hypothesis, $\widetilde{W}_i^{k}(\mathbf{S}) \cap S_i = \ddot{S}_i^{k}$. Thus, $s_i \in \ddot{S}_i^{k}$. 

We conclude that if $s_i$ is not NF-conditionally weakly dominated on $(\mathcal{S}_i, \widetilde{W}^k(\mathbf{S}))$, then there exists a relaxed belief system $\ddot{b}_i \in \ddot{B}_i^{k+1}$ such that for all $h_i \in H_i$, $s_i$ is rational at $h_i \in H_i$ for $\ddot{b}_i$, i.e., $s_i \in \ddot{S}_i^{k+1}$.\newline

\noindent ``$\supseteq$'': We next show that, if $s_i \in \ddot{S}_i^{k + 1}$, then $s_i \in \widetilde{W}_i^{k+1}(\mathbf{S}) \cap S_i$.

If $s_i \in \ddot{S}_i^{k+1}$, then there exists a relaxed belief system $\ddot{b}_i \in \ddot{B}_i^{k+1}$ such that for all $h_i \in H_i$ the strategy $s_i$ is rational at $h_i$ for $\ddot{b}_i(h_i)$. This implies that for all $h_i \in H_i$ such that $\ddot{S}_{-i}^{k, T_{h_i}} \cap S_{-i}^{T_{h_i}}(h_i) \neq \emptyset$, we have $supp(\ddot{b}_i(h_i)) = \ddot{S}_{-i}^{k, T_{h_i}}$. By the induction hypothesis, $\ddot{S}^{k, T_{h_i}}_{-i} = \widetilde{W}_{-i}^k(\mathbf{S}) \cap S_{-i}^{T_{h_i}}$. Hence, $\ddot{b}_i(h_i)$ has full support on $\widetilde{W}_{-i}^k(\mathbf{S}) \cap S_{-i}^{T_{h_i}}$.

If $s_i$ is rational at $h_i$ for $\ddot{b}_i(h_i)$, then for any $\tilde{s}_i \in S_i^{T_{h_i}}$ with $s_i \in [\tilde{s}_i]$, $\tilde{s}_i$ is rational at $h_i$ for $\ddot{b}_i(h_i)$. Thus, if $s_i$ is rational at $h_i$ for $\ddot{b}_i(h_i)$, then $\tilde{s}_i \in S_i^{T_{h_i}}$ with $s_i \in [\tilde{s}_i]$ is not weakly dominated in $\widetilde{W}_{-i}^k(\mathbf{S}) \cap S_{-i}^{T_{h_i}}$ because of Lemma 4 in Pearce (1984). It then follows that if  $s_i \in \ddot{S}_i^{k+1}$, then $s_i$ is not NF-conditionally weakly dominated on $(\mathcal{S}_i, \widetilde{W}^k(\mathbf{S}))$. Hence, $s_i \in \widetilde{W}_i^{k+1}(\mathbf{S}) \cap S_i$. This completes the proof of Proposition~\ref{IA-RPR}. \hfill $\Box$\\

Next, we show that prudent rationalizability is equivalent to relaxed prudent rationalizability.

\begin{prop}\label{PR-RPR} For every finite dynamic game with unawareness and every $i \in I$, $\bar{S}_{i}^{k} = \ddot{S}_{i}^{k}$, for all $k \geq 0$. Consequently, $\bar{S}_{i}^{\infty} = \ddot{S}_{i}^{\infty }$.
\end{prop}

\noindent \textbf{Proof of Proposition~\ref{PR-RPR}} The proof proceeds by induction. By definition, $\bar{S}_{i}^{0} = \ddot{S}_{i}^{0}$.

Assume $\bar{S}_{i}^{k} = \ddot{S}_{i}^{k}$. We show that $\bar{S}_{i}^{k+1} = \ddot{S}_{i}^{k+1}$.

``$\supseteq$'': Fix $s_i \in \ddot{S}_i^{k+1}$ and $h_i \in H_i$ such that $s_i$ allows for $h_i$ (otherwise $s_i$ is trivially rational at $h_i$). There exists a relaxed belief system $\ddot{b}_i \in \ddot{B}_i^{k+1}$ such that $s_i$ is rational at $h_i$. By definition of $\ddot{B}_i^{k+1}$, $\ddot{b}_i(h_i)$ has full support on $\ddot{S}_{-i}^{k, T_{h_i}}$. Assume $\ddot{S}_{-i}^{k, T_{h_i}} \cap S_{-i}^{T_{h_i}}(h_i) \neq \emptyset$. Then $\ddot{b}_i(h_i)\left( \ddot{S}_{-i}^{k, T_{h_i}} \cap S_{-i}^{T_{h_i}}(h_i)\right) > 0$. Thus, we can consider conditional probabilities $\ddot{b}_i(h_i)\left(\cdot \left| S_{-i}^{T_{h_i}}(h_i) \right. \right)$.  (Otherwise, if $\ddot{S}_{-i}^{k, T_{h_i}} \cap S_{-i}^{T_{h_i}}(h_i) = \emptyset$, then $s_i$ is trivially rational at $h_i$.)

Suppose by contradiction that $s_i$ is not rational at $h_i$ with $\ddot{b}_i(h_i)\left(\cdot \left| S_{-i}^{T_{h_i}}(h_i) \right. \right)$. There exists an $h_i$-replacement of $s_i$, call it $s_i / \tilde{s}_i^{h_i}$, such that $s_i / \tilde{s}_i^{h_i}$ yields a strictly higher expected payoff in $T_{h_i}$ given the belief $\ddot{b}_i(h_i)\left(\cdot \left| S_{-i}^{T_{h_i}}(h_i) \right. \right)$. That is, $u_i^{T_{h_i}}\left(s_i / \tilde{s}_i^{h_i}, \ddot{b}_i(h_i)\left(\cdot \left| S_{-i}^{T_{h_i}}(h_i) \right. \right) \right) > u_i^{T_{h_i}}\left(s_i, \ddot{b}_i(h_i)\left(\cdot \left| S_{-i}^{T_{h_i}}(h_i) \right. \right) \right)$. Since $u_i^{T_{h_i}}\left(s_i, \ddot{b}_i(h_i) \right) \geq u_i^{T_{h_i}}\left(s_i / \tilde{s}_i^{h_i}, \ddot{b}_i(h_i)\right)$, we must have $\ddot{b}_i(h_i)\left(S_{-i}^{T_{h_i}}(h_i)\right) < 1$. Now
\begin{eqnarray*} u_i^{T_{h_i}}\left(s_i, \ddot{b}_i(h_i) \right) & \geq & u_i^{T_{h_i}}\left(s_i / \tilde{s}_i^{h_i}, \ddot{b}_i(h_i)\right) \\
& = & \ddot{b}_i(h_i)\left(S_{-i}^{T_{h_i}}(h_i)\right) u_i^{T_{h_i}}\left(s_i / \tilde{s}_i^{h_i}, \ddot{b}_i(h_i)\left(\cdot \left| S_{-i}^{T_{h_i}}(h_i) \right. \right) \right) \\ & & + \left(1 - \ddot{b}_i(h_i)\left(S_{-i}^{T_{h_i}}(h_i)\right)\right) u_i^{T_{h_i}}\left(s_i / \tilde{s}_i^{h_i}, \ddot{b}_i(h_i)\left(\cdot \left| S_{-i}^{T_{h_i}}(h_i) \right. \right) \right) \\ & > & \ddot{b}_i(h_i)\left(S_{-i}^{T_{h_i}}(h_i)\right) u_i^{T_{h_i}}\left(s_i, \ddot{b}_i(h_i)\left(\cdot \left| S_{-i}^{T_{h_i}}(h_i) \right. \right) \right) \\ & & + \left(1 - \ddot{b}_i(h_i)\left(S_{-i}^{T_{h_i}}(h_i)\right)\right) u_i^{T_{h_i}}\left(s_i, \ddot{b}_i(h_i)\left(\cdot \left| S_{-i}^{T_{h_i}}(h_i) \right. \right) \right) \\ & =& u_i^{T_{h_i}}\left(s_i, \ddot{b}_i(h_i) \right),
\end{eqnarray*} a contradiction.

``$\subseteqq$'': Let $\bar{S}_{-i}^{k, T_{h_i}}(h_i)$ denote the set of opponents' level $k$-prudent rationalizable $T_{h_i}$-partial strategy profiles that allow for $h_i$. For each $T \in \mathbf{T}$ for which $H_i^T$ is nonempty, let $\bar{S}_{-i}^{k, T}(H_i^T) = \bigcup_{h_i \in H_i^T} \bar{S}_{-i}^{k, T}(h_i)$. Note that if $\bar{S}_{-i}^{k, T}(H_i^T) \subsetneqq \bar{S}_{-i}^{k, T}$, then there is a terminal history in $T$ that is excluded by any information set $h_i \in H_i^T$. (In general, $\bar{S}_{-i}^{k, T}(H_i^T)$ may not be a cross-product set).

We claim that there exists a nonempty subset of information sets $G^T_i \subseteq H_i^T$ such that $\left\{\bar{S}_{-i}^{k, T}(h_i)\right\}_{h_i \in G^T_i}$ forms a partition of $\bar{S}_{-i}^{k, T}(H_i^T)$. To define $G^T_i$, we first define the rank of an information set $h_i \in H_i^T$ as the maximal number of information sets in $H_i^T$ required to pass in order to allow for a terminal node in $T$. Note that, if $h_i \in H_i^T$ with $n \in h_i$ and there is path from $n$ to $n'$ in $T$ with $i \in I_{n'}$, then Property I3 (see Appendix~\ref{properties}) ensures that there exists an information set $h'_i \in H_i^T$ with $n' \in h'_i$. Note further that there may be terminal nodes in $T$ that are excluded by any information set in $H_i^T$.

Using the definition of rank of an information set, we construct a subset of information sets $G^T_i$ as follows: For any information set $h_i \in H_i^T$ of lowest rank, let $h_i \in G^T_i$ if there is no information set $h_i' \in H_i^T$ that precedes $h_i$. Otherwise, consider any information set $h_i' \in H_i^T$ of second lowest rank and let $h_i' \in G^T_i$ if there is no information set $h_i'' \in H_i^T$ that precedes $h_i'$, etc. Since $T$ is finite, $H_i^T$ is finite, and the procedure terminates after finite steps. (In particular, if player $i$ moves at the root of $T$, $G^T_i$ is a singleton whose only information set contains the root.)

By construction, $\bigcup_{h_i \in G^T_i} \bar{S}_{-i}^{k, T}(h_i)$ covers $\bar{S}_{-i}^{k, T}(H_i^T)$. Moreover, from Property I6 (see Appendix~\ref{properties}) follows that for any two $h_i, h_i' \in G^T_i$ with $h_i \neq h_i'$ there is no profile of opponents' $T$-partial strategies $s_{-i}^T \in \bar{S}_{-i}^{k, T}(h_i) \cap \bar{S}_{-i}^{k, T}(h_i')$.

Fix $s_i \in \bar{S}_i^{k + 1}$ and $h_i \in H_i$ such that $s_i$ allows for $h_i$ (otherwise $s_i$ is trivially rational at $h_i$). There exists a belief system $\bar{b}_i \in \bar{B}_i^{k + 1}$ such that $s_i$ is rational at $h_i$. By definition of $\bar{B}_i^{k + 1}$, the support of $\bar{b}_i(h_i)$ is the set $\bar{S}_{-i}^{k, T_{h_i}}(h_i)$.

Construct a belief system $\ddot{b}_i$ by setting for arbitrary $\varepsilon \in (0, 1)$,
\begin{eqnarray*} \ddot{b}_i(h_i)(s_{-i}) & := & \left\{ \begin{array}{cl} \frac{1 - \varepsilon}{\left|G^{T_{h_i}}_i \right|} \bar{b}_i(h_i)(s_{-i}) & \mbox{ if } s_{-i}^T \in \bar{S}_{-i}^{k, T_{h_i}}(h_i), \mbox{ and } \\  \frac{\varepsilon}{\left| \bar{S}_{-i}^{k, T_{h_i}} \setminus \bar{S}_{-i}^{k, T_{h_i}}(H_i^{T_{h_i}}) \right|} & \mbox{ if } s_{-i}^T \in \bar{S}_{-i}^{k, T_{h_i}} \setminus \bar{S}_{-i}^{k, T_{h_i}}(H_i^{T_{h_i}}). \end{array} \right.
\end{eqnarray*} This defines a probability measure $\ddot{b}_i(h_i)$ with full support on $\bar{S}_{-i}^{k, T_{h_i}}$, for each information set $h_i \in H_i$.

Let $\ddot{B}_i^{k + 1}$ be the set of all belief systems defined as above from any $\bar{b}_i \in \bar{B}_i^{k + 1}$ and $\varepsilon \in (0, 1)$.

Finally, note that since for any strategy $s_i \in \bar{S}_i^{k + 1}$ there exists some $\bar{b}_i \in \bar{B}_i^{k + 1}$ such that $s_i$ is rational at every information set $h_i \in H_i$, $s_i$ continues to be rational with a belief system $\ddot{b}_i \in \ddot{B}_i^{k + 1}$ at every information set $h_i \in H_i$. Thus $s_i \in \ddot{S}_i^{k + 1}$. This completes the proof of Proposition~\ref{PR-RPR}\hfill $\Box$\\

The proof of the theorem now follows from Propositions~\ref{IA-RPR} and~\ref{PR-RPR}. \hfill $\Box$



\end{document}